\documentclass[useAMS,usenatbib,twocolumn]{mnras}
\usepackage{times}
\usepackage{url}
\usepackage[dvips]{graphicx}
\DeclareGraphicsExtensions{.pdf,.png,.jpg,.mps,.eps,.ps}
\usepackage{multicol}
\usepackage{subfig}
\usepackage{amsmath, array}
\usepackage{wrapfig}

\newcommand\asca{{\it ASCA}}
\newcommand\swift{{\it Swift}}

\newcommand\rosat{{\it ROSAT}}

\newcommand\xmm{{\it XMM--Newton}}
\newcommand\flux{\ifmmode {\rm~erg cm}$^{-2}$\ ; {\rm s}$^{-1}$ \else ~erg cm$^{-2}$ s$^{-1}$\fi}
\newcommand\kms{\ifmmode {\rm~km\ s}$^{-1}$ \else ~km s$^{-1}$\fi}
\newcommand\Hunit{\ifmmode {\rm~km\ s}$^{-1}$\ {\rm Mpc}$^{-1}$
        \else ~km s$^{-1}$ Mpc$^{-1}$\fi}
\newcommand\ctssec{\ifmmode {\rm~count\ s}$^{-1}$ \else ~count s$^{-1}$\fi}
\newcommand\ergsec{\ifmmode {\rm~erg\ s}$^{-1}$ \else
        ~erg s$^{-1}$\fi}
\newcommand\funit{\ifmmode {\rm~erg\ s}$^{-1}$\ ; {\rm cm}$^{-2}$ \else
        ~ergs s$^{-1}$ cm$^{-2}$\fi}
\newcommand\phflux{\ifmmode {\rm~photon\ s}$^{-1}$\  ; {\rm cm}$^{-2}$
        \else   ~photon s$^{-1}$ cm$^{-2}$\fi}
\newcommand\efluxA{\ifmmode {\rm~erg\ s}$^{-1}$\ ; {\rm cm}$^{-2}$\ ; {\rm
        \AA}$^{-1}$ \else ~erg s$^{-1}$ cm$^{-2}$ \AA$^{-1}$\fi}
\newcommand\efluxHz{\ifmmode {\rm~erg\ s}$^{-1}$\ ; {\rm cm}$^{-2}$\ ; {\rm
        Hz}$^{-1}$ \else ~erg s$^{-1}$ cm$^{-2}$ Hz$^{-1}$\fi}
\newcommand\cc{\ifmmode {\rm~cm}$^{-3}$ \else cm$^{-3}$\fi}
\newcommand\fwhm{\ifmmode {\rm~FWHM} \else ${\rm~FWHM}$\fi}
\newcommand\msun{\ifmmode M_{\odot} \else $M_{\odot}$\fi}
\newcommand\lsun{\ifmmode L_{\odot} \else $L_{\odot}$\fi}

\newcommand\hbeta{\ifmmode {\rm H}\beta \else H$\beta$\fi}
\newcommand\kalpha{\ifmmode {\rm K}\alpha \else $K_\alpha$\fi}
\newcommand\lalpha{\ifmmode {\rm L}\alpha \else $L_\alpha$\fi}
\newcommand\nh{\ifmmode N_{\rm H} \else N$_{\rm H}$\fi}
\newcommand\hhh{1H~0707--495}
\title{Complex UV/X--ray variability of 1H~0707--495}
\author[Pawar et al]{
P. K. Pawar$^{1}$\thanks{PKP:pawar.pk123@gmail.com}, G. C. Dewangan$^{2}$, I. E. Papadakis$^{3,4}$, M. K. Patil$^{1}$, 
 Main Pal$^{2}$, A. K. Kembhavi$^{2}$\\
$^{1}$Swami Ramanand Teerth Marathwada University, Nanded--431 606, India\\
$^{2}$Inter University Center for Astronomy and Astrophysics, Pune--411 007, India\\
$^{3}$Department of Physics and Institute of Theoretical and Computational Physics, University of Crete, 71003, Heraklion, Greece\\
$^{4}$IESL, Foundation of Research and Technology, 71110 Heraklion, Greece}

\date{Accepted XXX. Received YYY; in original form ZZZ}

\pubyear{2017}

\begin{document}
\label{firstpage}
\pagerange{\pageref{firstpage}--\pageref{lastpage}}
\maketitle

\begin{abstract}
We study the relationship between the UV and X--ray variability of the narrow--line Seyfert 1 galaxy \hhh. Using a year long \swift{} monitoring and four long \xmm{} observations, we perform cross--correlation analyses of the UV and X--ray light curves, on both long and short time scales. We also perform time--resolved X--ray spectroscopy on 1--2 ks scale, and study the relationship between the UV emission and the X--ray spectral components -- soft X--ray excess and a power--law. We find that the UV and X--ray variations anti--correlate on short, and possibly on long time scales as well. Our results rule out reprocessing as the dominant mechanism for the UV variability, as well as the inward propagating fluctuations in the accretion rate. Absence of a positive correlation between the photon index and the UV flux suggests that the observed UV emission is unlikely to be the seed photons for the thermal Comptonisation. We find a strong correlation between the continuum flux and the soft--excess temperature which implies that the soft excess is most likely the reprocessed X--ray emission in the inner accretion disc. Strong X--ray heating of the innermost regions in the disc, due to gravitational light bending, appears to be an important effect in \hhh, giving rise to a significant fraction of the soft excess as reprocessed thermal emission. We also find indications for a non-static, dynamic X--ray corona, where either the size or height (or both) vary with time.

\end{abstract}

\begin{keywords}
 galaxies: active - galaxies: individual : {\hhh} - galaxies: Seyfert
 - X--rays: galaxies
\end{keywords}

%====================================
\section{Introduction}
%====================================
The primary continuum of radio-quiet active galactic nuclei (AGN) consists of the optical/UV big blue bump (BBB) and the X--ray power law with an energy cutoff at $\sim100$~keV (e.g. \citealt{2014ApJ...782L..25M};\citealt{2016MNRAS.458.2454L}). The BBB can extend from $1\mu m$ to $1000{\rm~\AA}$, and sometimes into the ``soft'' X--ray band, and is thought to be the thermal emission from an optically thick accretion disk. It is generally accepted that the X--ray power law arises from the Compton up-scattering of low energy disk photons by the hot electron cloud present in the Comptonizing corona (e.g. \citealt{1976ApJ...204..187S}; \citealt{1985ApJ...289..514Z}; \citealt{1980A&A....86..121S}; \citealt{1991ApJ...380L..51H}). However, the geometry and origin of the corona is still unclear. 

On top of the X--ray power law several other features are also observed e.g., the reflection hump in the ~\rm{20 -- 50}~keV, the (often broad) Fe \kalpha{} fluorescent line at $\sim 6.4$~keV, and the soft X--ray excess (SE) emission below $2$~keV which is seen as emission in excess of the $2-10$~keV power law fit, when extrapolated at lower energies. The origin of the reflection hump and of the Fe \kalpha{} line is attributed to the reprocessed emission (also commonly referred as reflection) of the power law photons by the material in the vicinity of the central source, either the disc or the obscuring torus \citep{1991MNRAS.249..352G, 2002MNRAS.331L..35F}. However, the origin of the SE is less certain even three decades after its discovery \citep{1985ApJ...297..633S, 2005AIPC..774..317S, 2007ApJ...671.1284D}. Two of the most popular and currently accepted ideas are blurred ionized reflection from the inner part of the disk \citep{2005MNRAS.358..211R} and the Comptonization model (\citealt{2013A&A...549A..73P,2012MNRAS.420.1848D}). Both the reflection and Comptonization models fit reasonably well this component. Therefore, it is difficult to differentiate between the two models just by spectral studies. 

Additional information is provided by the large amplitude, fast variations that AGNs exhibit at all wavelengths. Simultaneous optical/UV and X--ray variability studies, in particular, can provide important information on the relationship between the variations observed in these bands, which can then be compared with the different model predictions. For example, the reflection model assumes that the cold disk and/or torus is illuminated by the primary X--ray photon source. Part of the X--ray illuminating flux is expected to be absorbed and result in extra optical/UV emission, which will add to the observed BBB. If the primary X--ray emission is variable, the reprocessed component should also vary, and we should observe the X--rays variations to drive the variations in optical/UV band with a delay. Observationally, such lags have been detected in several sources (e.g. NGC4051: \citealt{2002ApJ...580L.117M}, ARK564: \citealt{2001ApJ...561..162S}, NGC2617: \citealt{2014ApJ...788...48S}, NGC5548: \citealt{2014MNRAS.444.1469M} and \citealt{2015ApJ...806..129E}). In all cases though, the correlation between the X--ray and the optical/UV variations is moderate/weak. Recently, \cite{2017MNRAS.464.3194B} reported the results from a systematic study of the simultaneous UV/X--ray variations in 21 AGN, using archival {\it Swift} data. They found (weak) X--ray/UV correlated variability in less than half of the objects in their sample. In all cases, the UV variations were lagging those observed in the X--ray band. 

In the case of the Comptonization model, the SE is thought to be produced by thermal Comptonization of disk photons by a low temperature, optically thick corona. This model essentially requires two distinct coronae -- one to produce the primary X--ray photons and another one to produce the SE. It is not obvious what should be the correlation between the optical/UV and the X--ray variations in this case. If accretion rate variations propagate from the outer to the inner part of the disc and/or the SE photons are the input photons to the hard X--ray emitting corona, then a delay of the X--ray variations, with respect to the optical/UV variations, should be expected. There are a few sources where such lag has been observed (e.g. MCG--6--3--0--15: \citealt{2005A&A...430..435A} and PKS~0558--504: \citealt{2013MNRAS.433.1709G}). However, in a few other similar sources no such lag has been observed (MRK335: \citealt{2012ApJS..199...28G}). 

Multi--wavelength variability studies can play an important role in order to understand the physical mechanism that drives the emission variations in different bands and, thus, clarify the complex interaction between the optical/UV, soft, and hard X--ray emission. Such studies can be performed with the use of data from current X--ray space missions such as \xmm{}, \swift{} \& \emph{ASTROSAT} which can provide simultaneous X--ray and UV/Optical band observations. We present the results from a detailed study of the short as well as the long--term UV and X--ray variability of \hhh, using all the available \swift{} and \xmm{} observations. 

\hhh{} (\emph{z} = 0.0411) is a Narrow Line Seyfert 1 (NLS1), which hosts a relatively low mass ($M_{BH}\sim 5.6\times10^6M_{\odot}$; \citealt{2016ApJ...819L..19P}). It was initially detected by \citet{1986ApJ...301..742R} using the \emph{HEAO~I} scanning modulation collimator. Its optical spectrum shows broad hydrogen lines (FWHM of \hbeta{} = 1000 {\rm~km s}$^{-1}$), very strong Fe II lines, and weak forbidden lines. Its soft X--ray emission properties and its variability were studied using \rosat{} and \asca{} \citep{1996A&A...305...53B, 1999ApJ...524..667T, 1999ApJS..125..297L}. 
\hhh{} received significant attention after its first observation by \xmm{}, where it exhibited a sharp spectral drop at $\sim$ 7~keV and a strong variability by a factor of $\sim$ 4. The flux drop was assumed to be due either to partial--covering absorption by neutral gas cloud or to strong reflection of X--ray photons from an ionized disc \citep{2002MNRAS.329L...1B}. Both interpretations require iron overabundance to explain the observed drop in flux. The overabundance of iron was expected to lead to accompanying iron L--line complexes, which were later detected by \citet{2009Natur.459..540F}. The same authors also observed a reverberation lag of about $30$~s between the direct X--ray continuum and the soft band. The object was observed more times with \xmm{} and the resulting data have been studied by several researchers in order to understand the detailed spectral behaviour and origin of the reverberation lag (e.g. \citealt{2011MNRAS.414.1269W, 2012MNRAS.424.1284W, 2012MNRAS.422.1914D, 2013MNRAS.428.2795K, 2016A&A...594A..71E}). 

The detection of iron L and K lines and the soft/hard X--ray negative lag of $\sim 30$~s suggest strong X--ray illumination of the innermost regions of the disc. This strong illumination may also result in optical/UV reprocessed emission. Since \hhh{} has been extensively observed by both \xmm{} and \swift{} the last decade, it is a good candidate to study the complex interaction between the disc and the X--ray corona through a correlation analysis of the archival UV and X--ray data from these two satellites. Recently, \cite{2015MNRAS.453.3455R} published the results from a correlation study of the source, using the archival \xmm{} observations. They reported a weak correlation between the X--rays and UV variations, with the later leading the X--ray variations by about $\sim$6.4 day. The lag magnitude is comparable to the total length of the observations (which casts some doubt on its significance). Nevertheless, Monte--Carlo simulations revealed that this delay was (moderately) significant at the $\sim 95$\% level. 

 We investigate the short and long term UV/X--ray correlation, and we compliment the flux correlation study with time resolved spectroscopy (using the \xmm{} data only) and a correlation investigation between the resulting spectral parameters, on short time scales. Our results reveal an unexpected $anti-$correlation between the UV and X--ray variations on long time scales, and no correlation at short time scales. These results rule out X--ray reprocessing, or inward propagating mass accretion rate fluctuations, as possible causes for the UV variations. In addition, the time resolved spectroscopy results strongly support the hypothesis of strong X--ray reprocessing in the inner disc. If both the inner disc and the X--ray source are close to the central black hole, strong relativistic effects will prevent the X--rays to illuminate the outer disc, thus explaining the lack of (positive) correlation between the observed X--ray and UV variations. 
 
 %====================================
\section{Observations and data reduction}
%====================================
%====================================
\subsection{\swift{} monitoring}
%====================================
\hhh{} was monitored by the \swift{} observatory \citep{2004ApJ...611.1005G} for a year, between April 2010 and March 2011 with
an interval of $\sim3-4$ days between consecutive observations. During this year long monitoring, the object was observed daily during 18 -- 31 January, 2011 when it was in a low flux state, which actually lasted for almost two months (between December 2011 -- January 2012). During this low flux state \xmm{} also observed the source \citep{2012MNRAS.419..116F}. 
The \swift{} X--ray observations were made using the X--ray Telescope (XRT; \citealt{2005SSRv..120..165B}) while the UV observations were made with the UV and Optical Telescope (UVOT; \citealt{2005SSRv..120...95R}) with the UVW2 filter. The central wavelength of UVW2 filter is 1928{\rm~\AA}, and it covers the wavelength range between 1800 -- 2600{\rm~\AA}.

We performed photometry using the UVW2 images acquired with the UVOT. We followed the \swift{} UVOT Software Guide
\footnote{\url{http://swift.gsfc.nasa.gov/analysis/UVOT\_swguide\_v2\_2.pdf}} for analyzing this dataset. The raw images were corrected for bad pixels, modulo--8 noise and flat fielding. Sky transformation was done using \emph{swiftxform} task. Resultant images were later used to perform the photometry using the task \emph{uvotsource}. A source region of 5\arcsec{} centred on the source and a nearby background region, devoid of any sources, of 15\arcsec{} was used to perform photometry. We filtered the UVOT data for possible `flux dropout' which occurs when the source falls on specific regions of the detector following the method given in the Appendix of \cite{2015ApJ...806..129E}. We could only find three dropout points which were subsequently removed from the further analysis.

The XRT light curves were derived in the soft (0.3 -- 1~keV) and hard (1.5 -- 3~keV) energy bands (SX and HX, respectively), using the ``Swift--XRT products'' generator web tool\footnote{\url{http://www.swift.ac.uk/user\_objects/}}. 
The energy limits were selected based on the results of \citet{2012MNRAS.422.1914D} and \citet{2009Natur.459..540F}, who showed that these bands are dominated by two different components; the hard band by the direct, continuum emission and the soft band by the soft--excess emission (which may be due to X--ray reflection). The web tool generates the image, light curves and spectra, if the source counts are sufficient using all the available datasets. Light curves in the same bands were also generated manually using the \emph{XSELECT} for cross verification. The results were consistent and no significant deviations were observed. Therefore, we prefer to use the pipeline results as those incorporate additional finer steps like pile--up correction.

%%%%%%%%%%%%%%%%%%%%%%
% Fig. 1
%%%%%%%%%%%%%%%%%%%%%%
\begin{figure}
  \centering
  \includegraphics[scale=0.33]{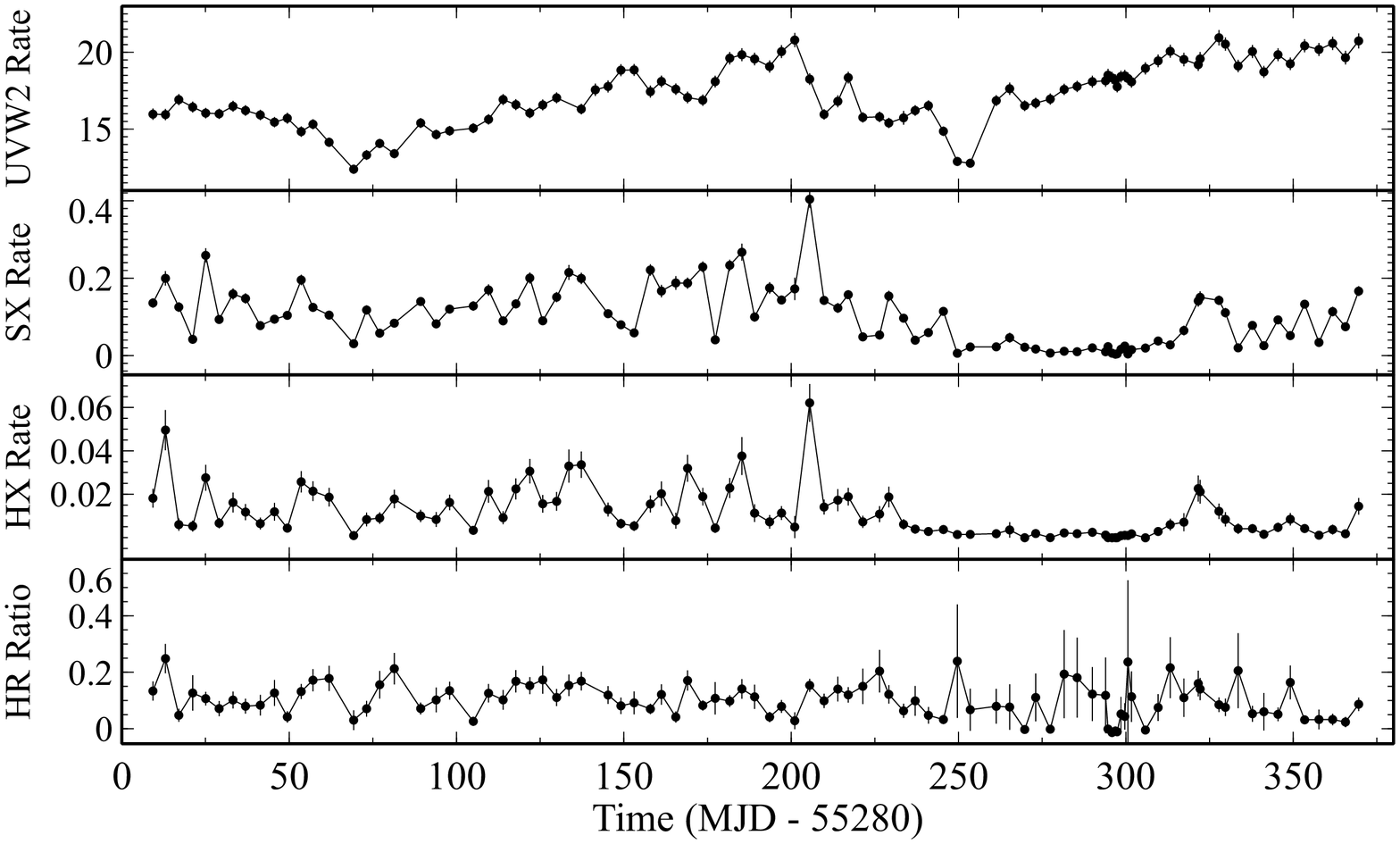}
  \includegraphics[width=8.3cm,height=3cm]{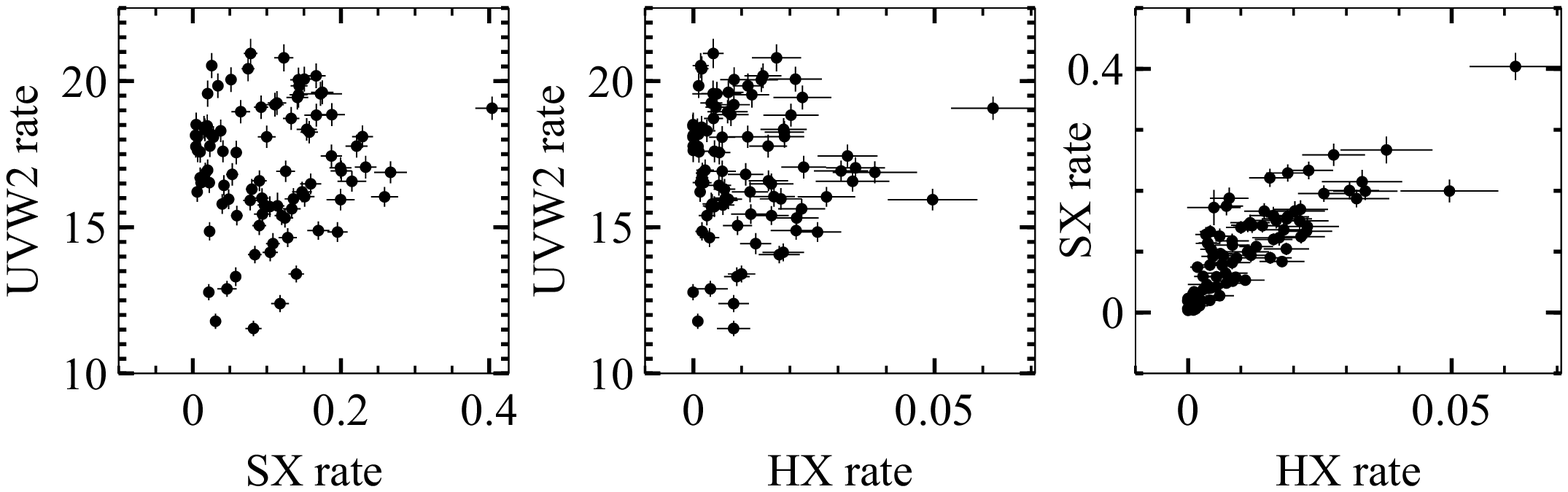}
  \caption{\emph{Top:} The UVW2 ($\lambda{}_{eff}~\rm{\sim}$ 1920 \AA{}), SX (0.3 -- 1~keV) and HX (1.5 -- 3~keV) \swift{} light curves and the HR ratio evolution with time. \emph{Bottom:} Counts vs counts plots between different energy bands.}\label{swift_lc}
\end{figure}

Figure~\ref{swift_lc} shows the UVW2, SX AND HX long--term light curves over the whole period of the \swift{} observations. The bottom panel in the upper plot of the same figure shows the hardness ratio (HR), defined as the ratio of the HX over the SX count rate. The \swift{} light curves show clear X--ray and UV variations. To quantify this variability, we computed the fractional variability amplitude (F$_{var}$) of each light curve \citep{2003MNRAS.345.1271V}. The UVW2, SX and HX variability amplitudes are 12.4 $\pm{~\rm 0.2}$ \%, 76.7 $\pm{~\rm 1.1}$\% and 87.1 $\pm~\rm {3.5}$\%, respectively. Thus, the variability amplitude is energy dependent and decreases with decreasing energy. The hardness ratio also varies significantly. We get a $\chi^2$ value of 451 for 95 degrees of freedom (dof), when we fit a constant line to the data. This result indicates significant, long--term spectral variations, with a fractional variability amplitude F$_{var,\rm{HR}}=31.8\pm 5.1$, which is smaller than the two X--ray bands but larger than the UV variability amplitude.
 
Although the HX and SX light curves appear to be similar, this is not the case with the UVW2 light curve as well. The UV light curve shows broad, long term variations, which are not present in the X--ray light curves. The most striking example is the steady UV flux increase after $\sim 250$ days since the start of the observation, at the same time when the X--rays are in a very low state, and remain at a constant level, except perhaps the last $\sim 20$ days of the monitoring campaign, when the SX flux appears to increase as well. The lack of correlation between the UVW2 and the X--ray light curves is evident in the bottom panels of Fig.~\ref{swift_lc}, where we show the UVW2 count rate plotted as a function of the SX and HX. On the contrary, an excellent correlation is observed between the SX and HX count rates (bottom right panel of the same figure). 

%%%%%%%%% Table 1 %%%%%%%%%%
\begin{table*}
  \begin{center}
\caption{\xmm{} Observational details.}\label{obs_log}
  \begin{tabular}{c c c c c c}\hline
    Observation ID      &  Obs. Date 	  & Exposure (ks)& \# UVW1 Frames $^{a}$& Offaxis & Pileup\\\hline
    0511580101 (101)	& 2008-01-29 18:28:24 & 100	 & 67	& 0.012	& N \\
    0511580201 (201)	& 2008-01-31 18:20:46 & 80  	 & 61	& 0.012	& N \\
    0511580301 (301)	& 2008-02-02 18:22:11 & 87	 & 57	& 0.012	& N \\
    0511580401 (401)	& 2008-02-04 18:24:21 & 68	 & 50	& 0.012	& N \\\hline
  \end{tabular} 
  \end{center} 
 $^{a}$ Useful UVW1 frames after filtering.
\end{table*}

%====================================
\subsection{\xmm{} observations}\label{xmm_analysis}
%====================================
\hhh{} was observed with \xmm{} \citep{2001A&A...365L...1J} on various occasions. Continuous observations, over many orbits, have only been carried out during 29 January -- 6 February 2008 and 13--19 September 2010. During the 2008 observations the Optical Monitor (OM; \citealt{2001A&A...365L..36M}) was mostly operated in the imaging mode with the UVW1 filter along with European Photon Imaging Camera (EPIC--pn, \citealt{2001A&A...365L..18S}). There are more than 300 frames taken with this filter, resulting in a data set which is ideal for the study of the X--ray/UV correlation on times scales as short as a few hundred seconds (as opposed to a few days, in the case of the \swift{} data). The typical exposure time of the OM observations is $\sim 2$~ks. During the 2010 observations, the OM was operated with various filters. Furthermore, the total number of OM observations is much smaller. Consequently, it is not possible to probe the short timescale X--ray/UV correlation with this data set. In addition to these observations, the source was also observed once in 2000 and four times in 2007 for $\sim{}~40$~ks. These observations are too short to be useful in this study.

We acquired the Observation Data Files (ODFs) for the 2008 \xmm{} observations from the \emph{HEASARC} \footnote{\url{http://heasarc.gsfc.nasa.gov/cgi-bin/W3Browse/w3browse.pl}} archive. The detailed observation log of \hhh{} is given in Table ~\ref{obs_log}. We used both the OM and EPIC--PN data. The data were reprocessed and filtered using the \xmm{} Science Analysis System SASv13.5\footnote{\url{http://xmm.esac.esa.int/external/xmm\_data\_analysis/}} with the latest calibration files and following the SAS ABC guide\footnote{\url{http://xmm\_newton.abc-guide}}. We used the standard SAS processing script \emph{omichain} to reduce the OM imaging data. We reprocessed all the datasets using the script \emph{epproc} to produce calibrated X--ray images. We considered good quality, single and double events by setting \emph{FLAG} = 0 and \emph{PATTERN} $\leq$ 4. By generating background light curves above 10~keV, each dataset was examined for the flaring particle background. The intervals where the data were severely affected by such background flaring were filtered appropriately. The observations were checked for possible pile--up but we found no evidence for pile up in any of them. 
%%%%%%%%%%%%%%%%%%%%%%
% Fig. 2
%%%%%%%%%%%%%%%%%%%%%%%
\begin{figure}
\centering
\includegraphics[scale=0.165]{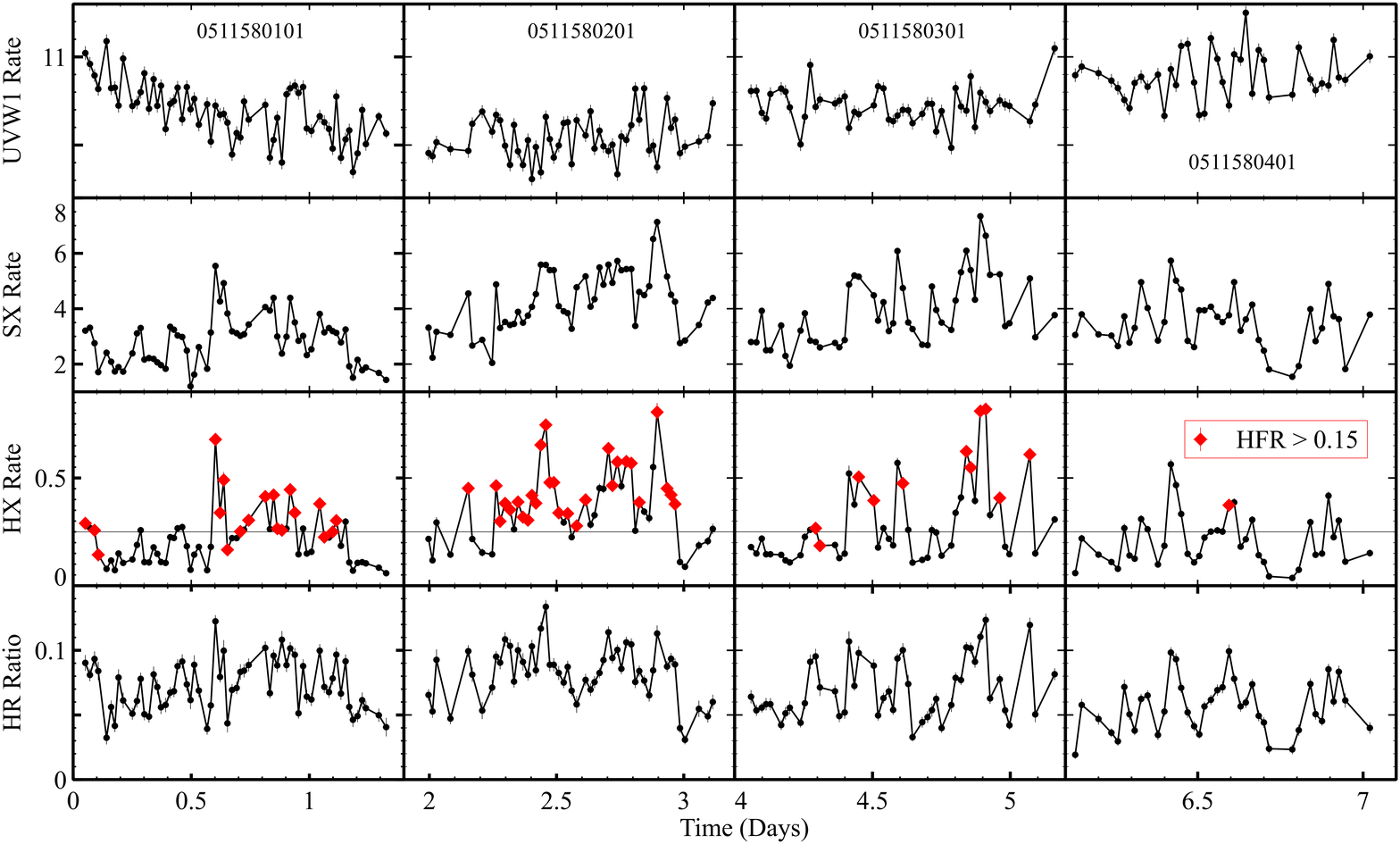}
\includegraphics[scale=0.42]{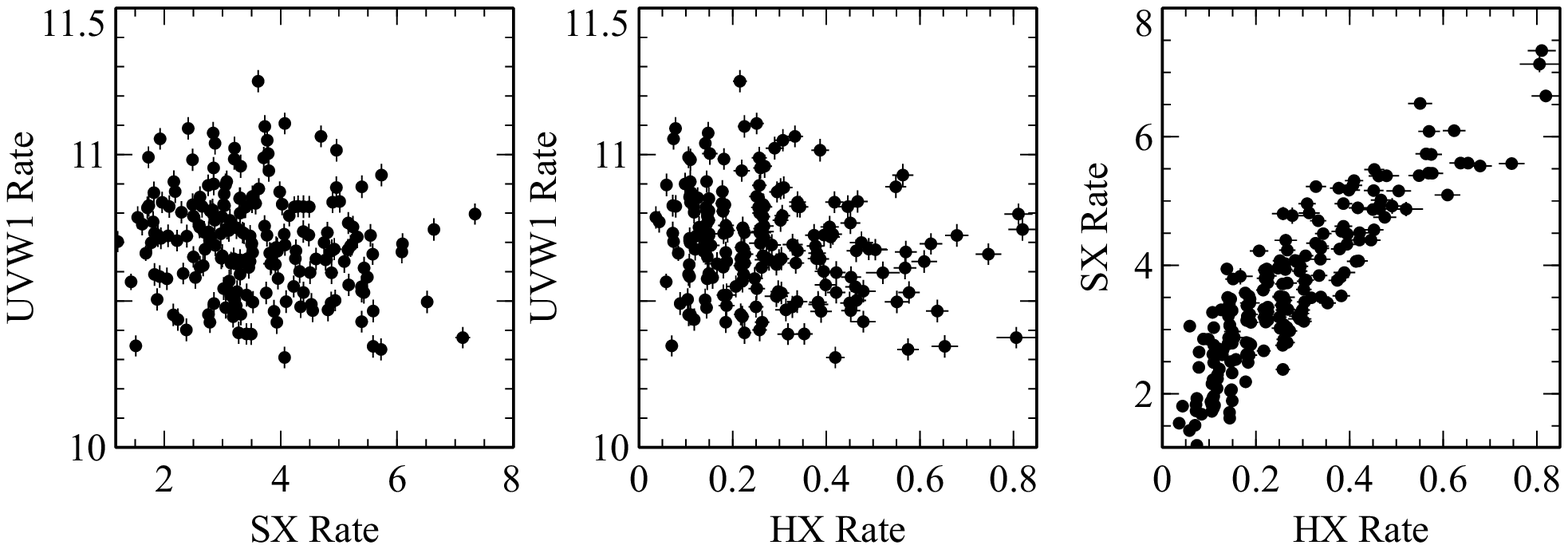}
\caption{\emph{Top:} The UVW1, soft (0.3--1~keV) and hard (1.5--3~keV) \xmm{} light curves. The fourth panel from the top shows the time evolution of the HR ratio (defined as the ratio of the hard over the soft X--ray band count rate). \emph{Bottom:} Count vs count plots between different energy bands.}\label{xmm_variability}
\end{figure}

We created time--selected event lists using the start and stop times of the UVW1 frames. In this way, we were able to construct light curves in the soft (0.3 -- 1~keV) and hard (1.5 -- 3~keV), X--ray bands, which are binned exactly like the OM light curves. Figure~\ref{xmm_variability} shows the resulting light curves. They clearly demonstrate significant X--ray and UV variations. The fractional variability amplitude of the HX, SX and UVW1 light curves is 58.6$\pm{0.5}$\%, 33.3$\pm{0.2}$\% and 1.6$\pm{0.1}$\%, respectively. The UVW1 variability amplitude is in agreement with that reported by ~\citep{2015MNRAS.453.3455R} using the same \xmm{} data. Just like with the long term variations, the short--term variability amplitude is energy dependent, increasing with increasing energy. We also computed the hardness ratio, as before (i.e. the ratio of the hard over the soft X--ray count rate). The HR is also variable (see fourth plot from top in Fig.~\ref{xmm_variability}), with an amplitude of F$_{var,{\rm HR}}=30.8 \pm 0.5$.

In the bottom panels of Fig.~\ref{xmm_variability} we plot the UVW2 count rate as a function of the SX and HX count rate. Similar to \swift{} observations, there is no apparent correlation between the observed UV and X--ray variations on short time scales as well. Contrary to this, the short--term soft and hard X--ray variations are positively correlated, to a high degree.

%====================================
\section{\xmm{} time resolved spectral analysis}\label{analysis_method}
%====================================
The \xmm{} X--ray count rate is much higher than the \swift{} count rate. Thus, we used the \xmm{} data to perform time resolved spectroscopy (TRS) to study the X--ray spectral evolution on short timescales. To this end, we used the final cleaned event lists and the time--selected event lists we mentioned in the previous section to extract the X--ray spectrum during the same time intervals of the OM frames. This allow us to perform correlation study between various spectral parameters and the UVW1 rate. We used a circular region of radius of 35$''$ to 40$''$ centred on the source. We subtracted background spectra which were extracted from source free circular regions on the same chip. The EPIC responses i.e., the redistribution matrix file (RMF) and the effective area file (ARF)
were generated using the SAS tasks \emph{rmfgen} and \emph{arfgen}, respectively. The resulting spectra were then grouped using {\it FTOOL} task \emph{grppha} to a minimum of 20 counts per spectral bin so that the $\chi^2$ minimization technique could be employed. 

%%%%%%%%%%%%%%%%%%%%%%%%%%%%%
% Fig. 3
\begin{figure*}
%%%%%%%%%%%%%%%%%%%%%%%%%%%%%
  \centering
  \includegraphics[scale=0.29]{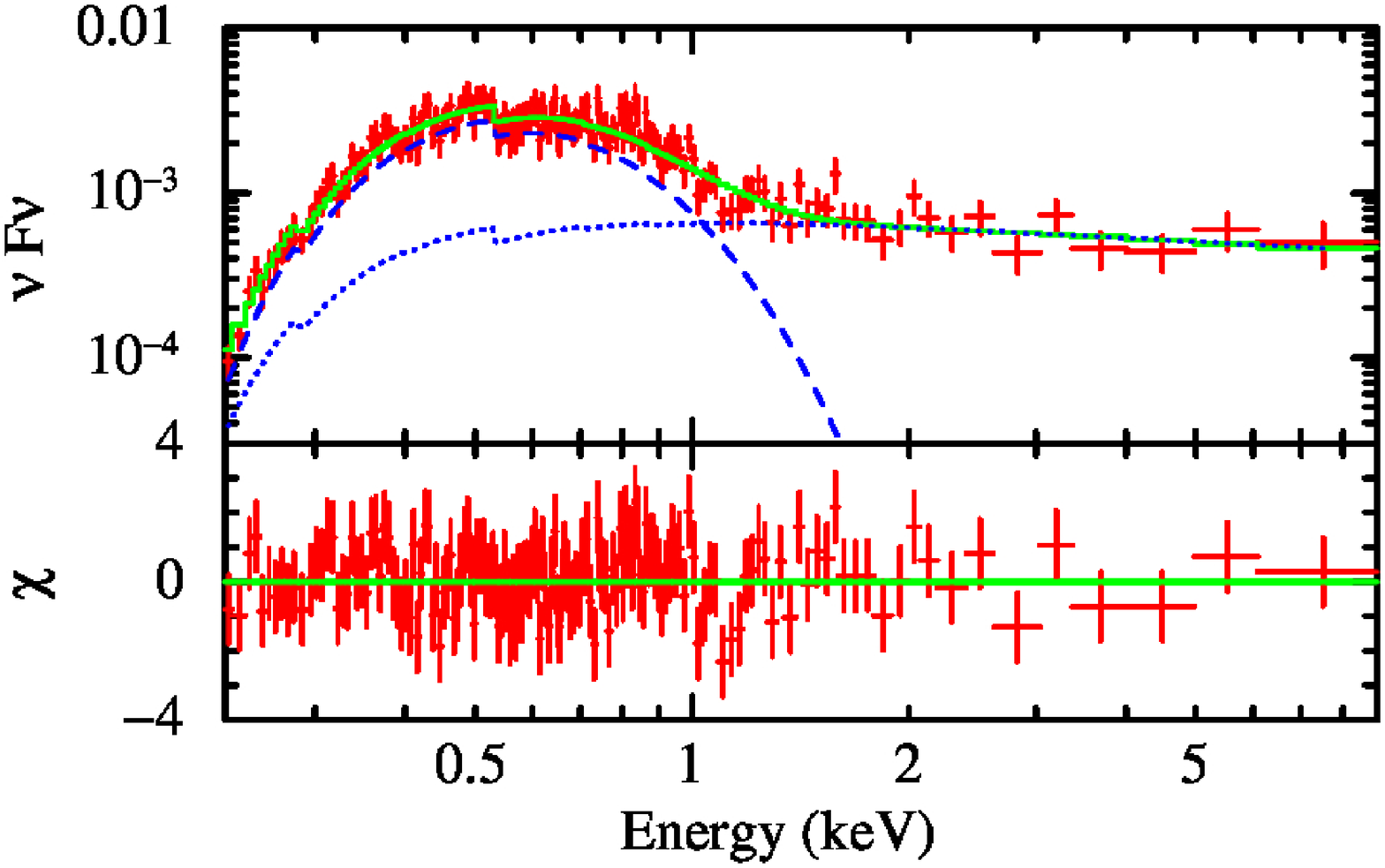}
  \includegraphics[height=4.3cm,width=6.5cm]{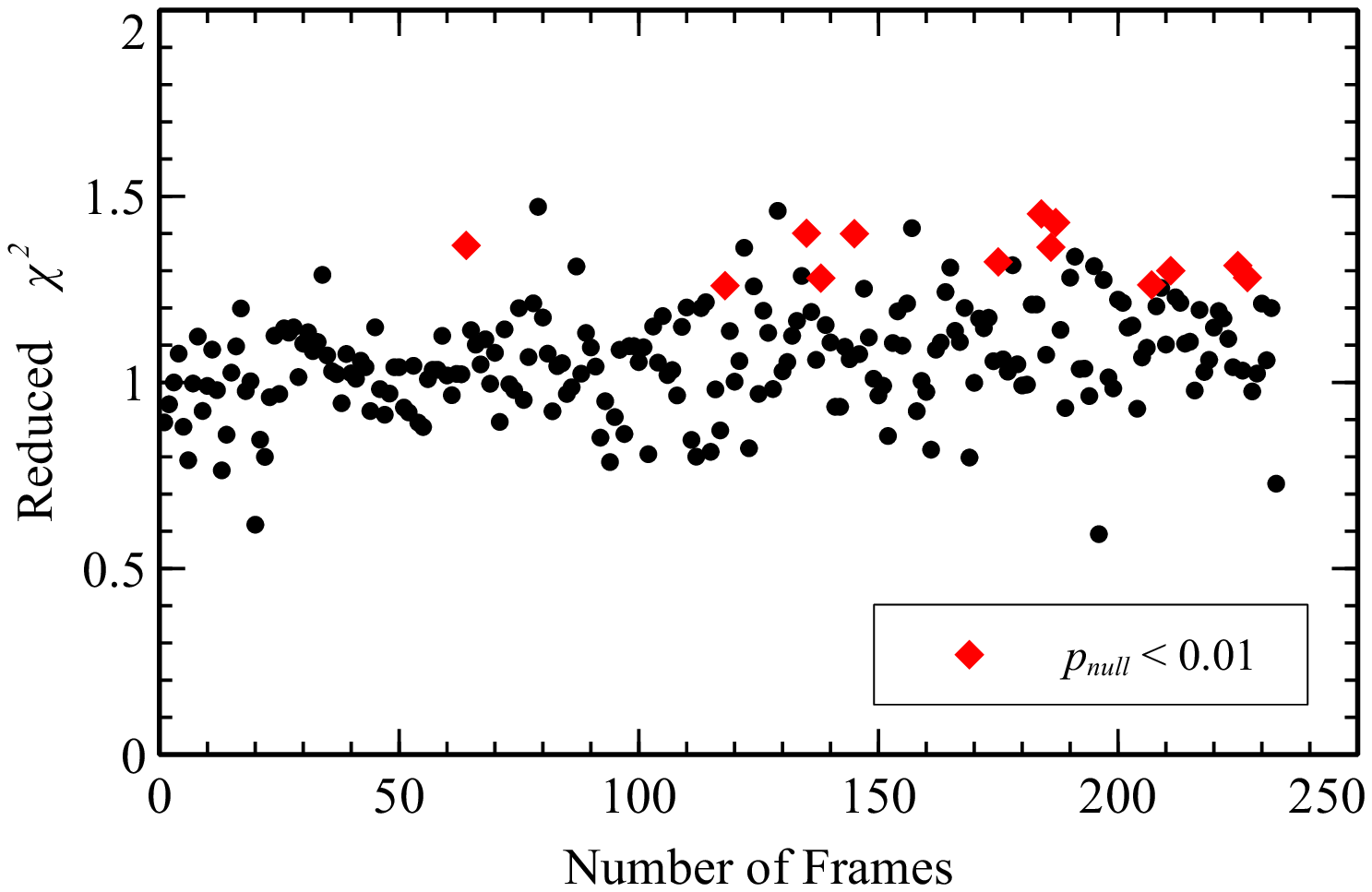}
  \caption{\emph{Left:} Typical \xmm{} spectrum and the respective best--fit models (blue dotted and dashed lines show the best--fit power law and black body component, respectively; the green continuous line shows the overall best--fit model). \emph{Right:} The best fit, $\chi^2$ reduced values for all spectra. Points in red indicate the spectra where the best--fit null hypothesis probability is less than 1\%.}\label{red_chi}
\end{figure*}

We fitted the time resolved spectra over the entire 0.3 -- 10~keV band with a simple, phenomenological model which consists of two components: a power law component (PL), to account for the continuum emission, and a black body component (BB) to account for the soft excess emission. Both the components were modified for Galactic absorption ($N_H=6\times10^{20}$ cm$^{-2}$ \citealt{1990ARA&A..28..215D}) by using the photo--electric absorption model $wabs$. Our adopted model is rather simple but, given the short duration of the OM exposures, the resulting X--ray spectra do not have the necessary quality to reveal all the complex features that have been seen in the past in the overall spectrum of the source. In any case, our simple model is capable to characterize the basic shape of the X--ray spectrum, as we show below. 

The left panel of Fig.~\ref{red_chi} shows a typical spectrum, the overall and the individual best--fit model components. The two components account for the emission in different regions of spectrum. The PL component accounts for the whole continuum emission above $\sim 2$~keV, while the black body accounts for the excess X--ray emission at lower energies (during the model fitting process, we always checked whether the PL and BB components are fitting the appropriate parts of the spectrum). The model fits the spectrum well ($\chi^2=142$ for $154$ degrees of freedom -- dof). The right~panel of Fig.~\ref{red_chi} shows the best fit,$\chi^2$ reduced values corresponding to the best--fit model of each spectrum. The fits are more or less acceptable. For example, there are just 13 spectra where the best--fit null hypothesis, $P_{null}$ (i.e. the model is consistent with the data) is smaller than 1\%. In these cases, the best--fit residuals show large deviations of the data from the best--fit model at a few energy points, which are distributed randomly over the whole energy range. We therefore conclude that the chosen model represents the overall shape of the observed spectra rather well. Finally, we computed the PL and BB (unabsorbed) flux ($f_{PL}$ and $f_{BB}$) in the energy bands 1.5--3 and 0.3--1~keV, respectively, using the {\tt XSPEC} convolution model $cflux$.

%====================================
\section{The TRS results}\label{TRSresults}
%====================================

%%%%%%%%%%%%%%%%%%%%%%%%%%%%%%%
% Fig. 4
%%%%%%%%%%%%%%%%%%%%%%%%%%%%%%%%
\begin{figure}
  \centering
  \includegraphics[scale=0.27]{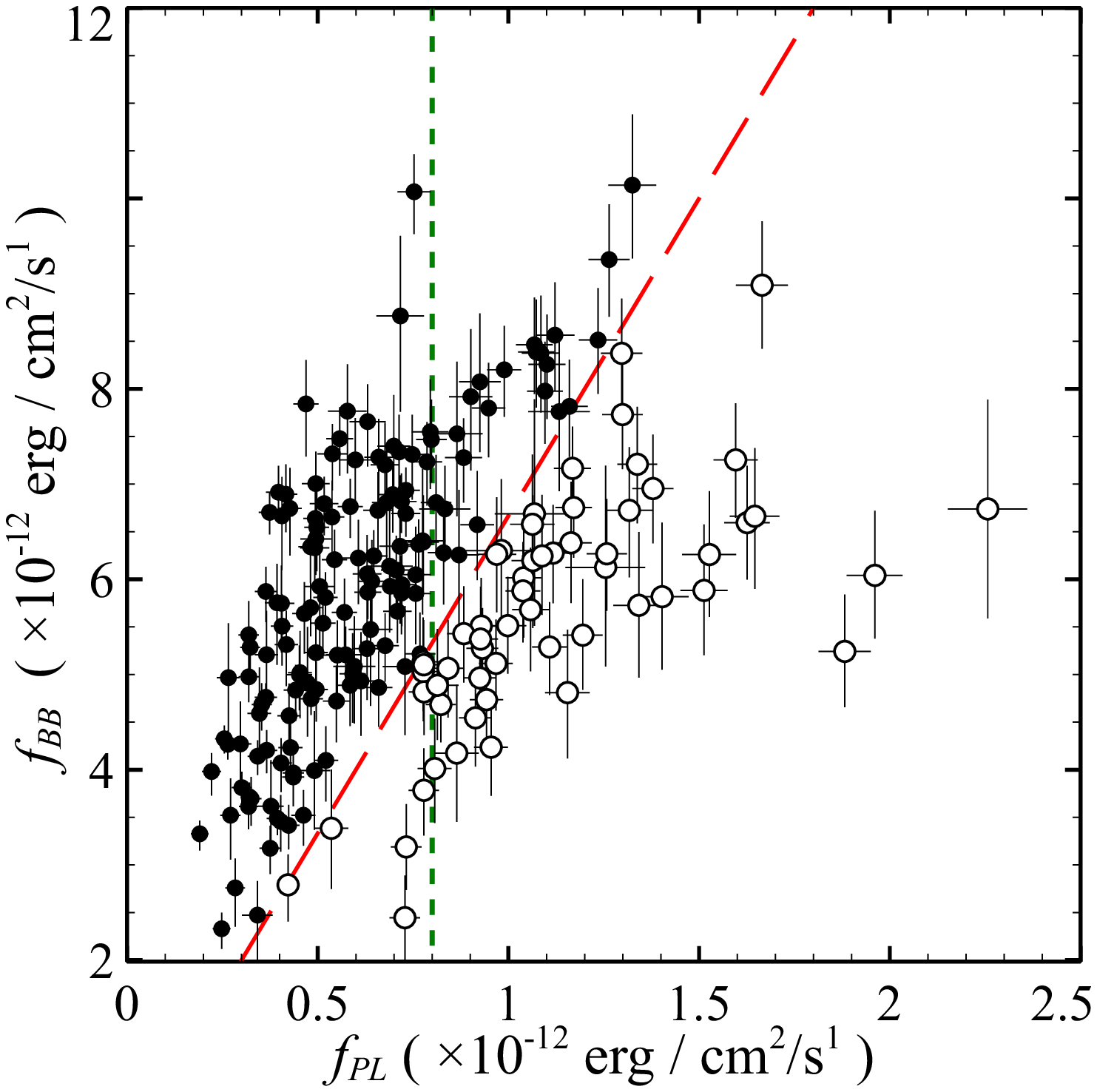}
  \includegraphics[scale=0.27]{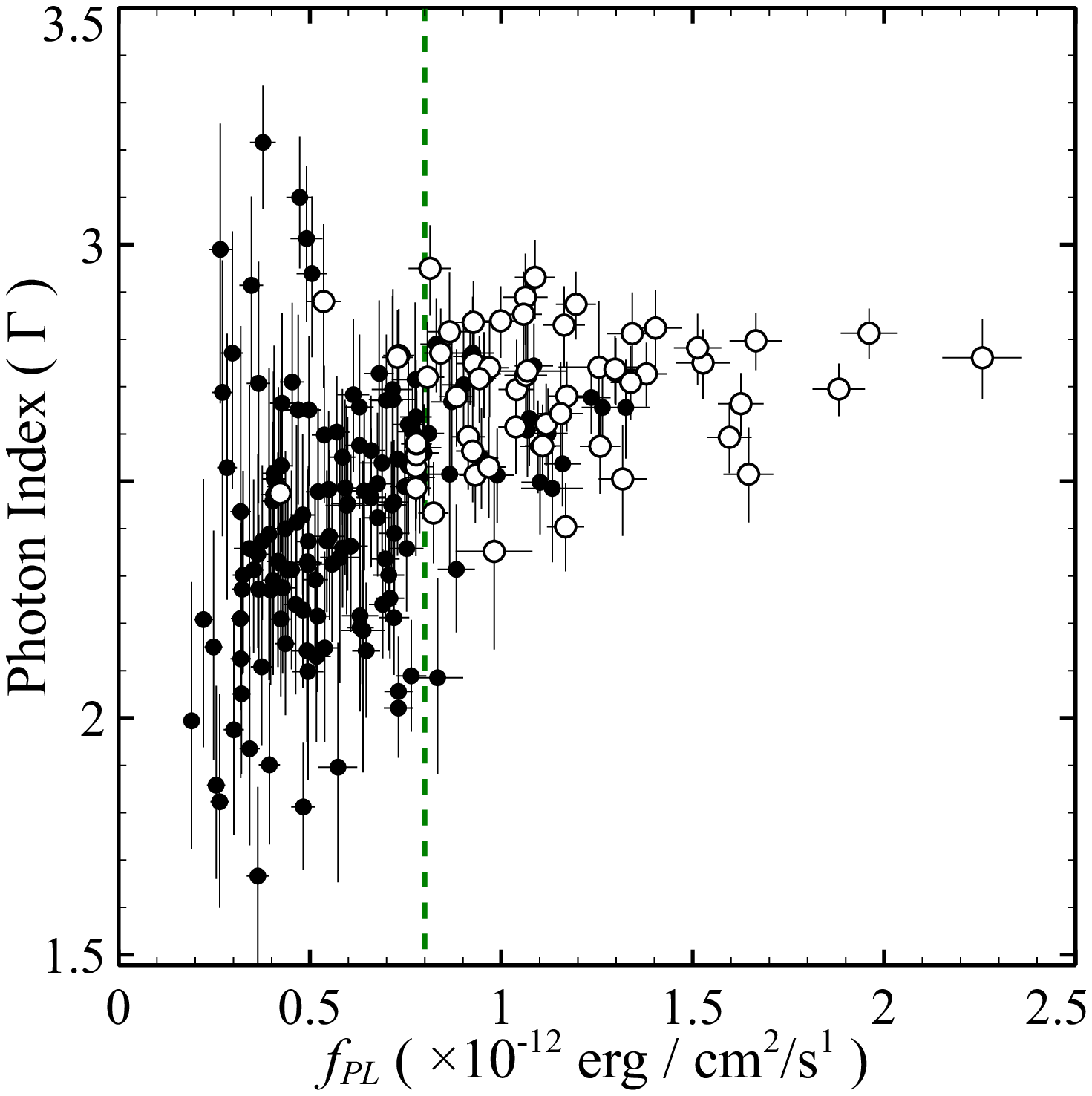}
  \includegraphics[scale=0.27]{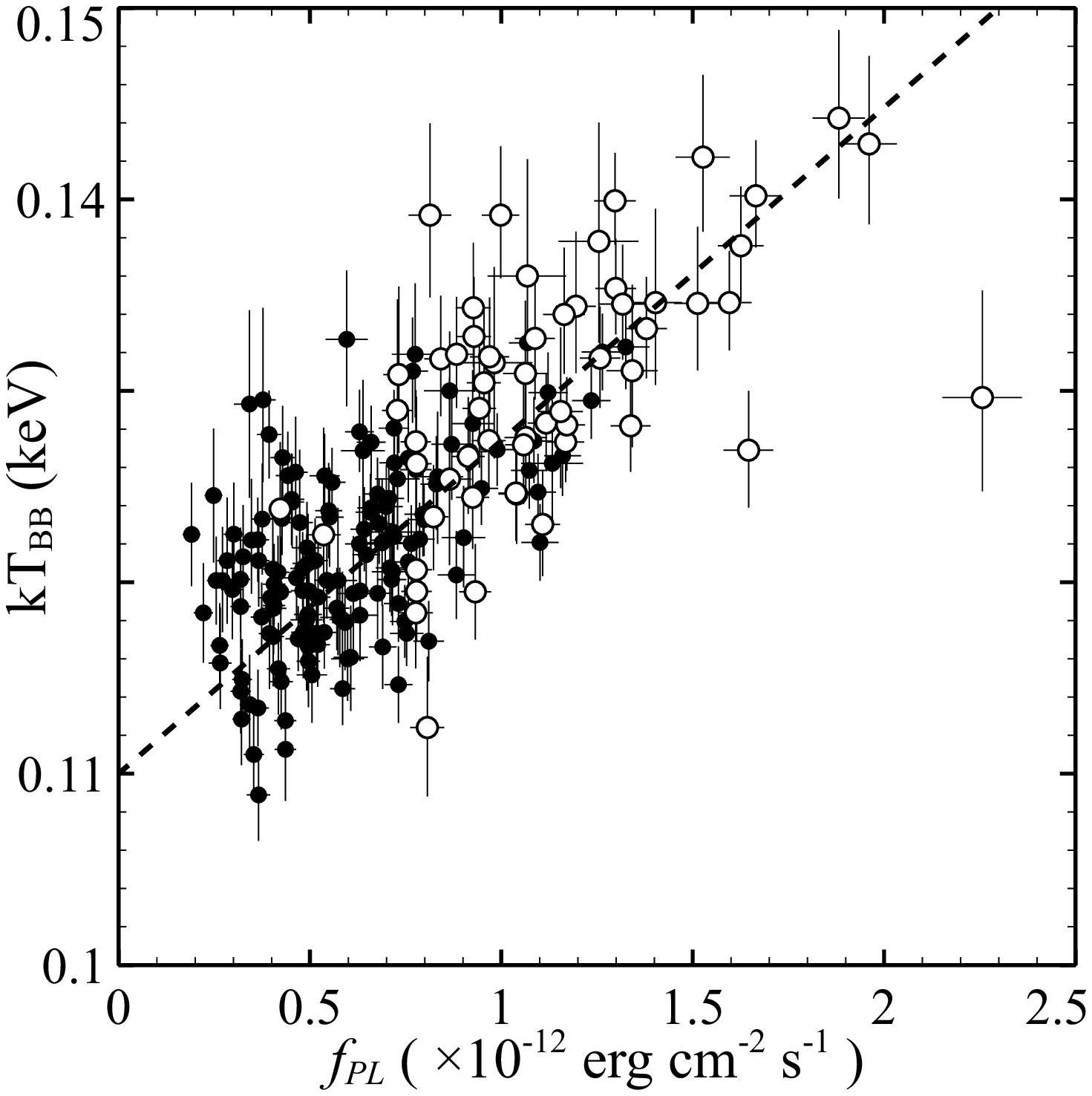}
  \includegraphics[scale=0.27]{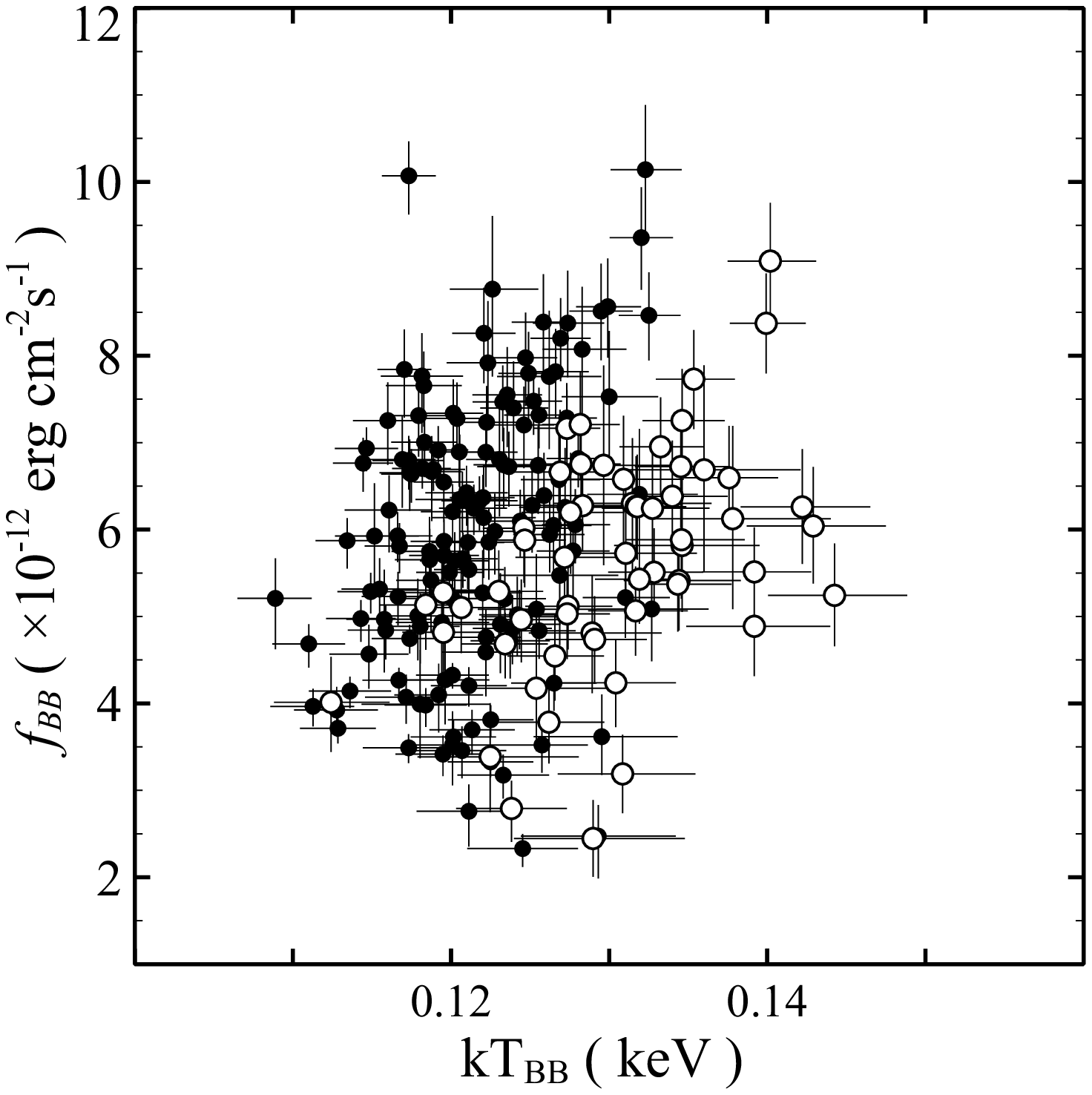}
  \includegraphics[scale=0.27]{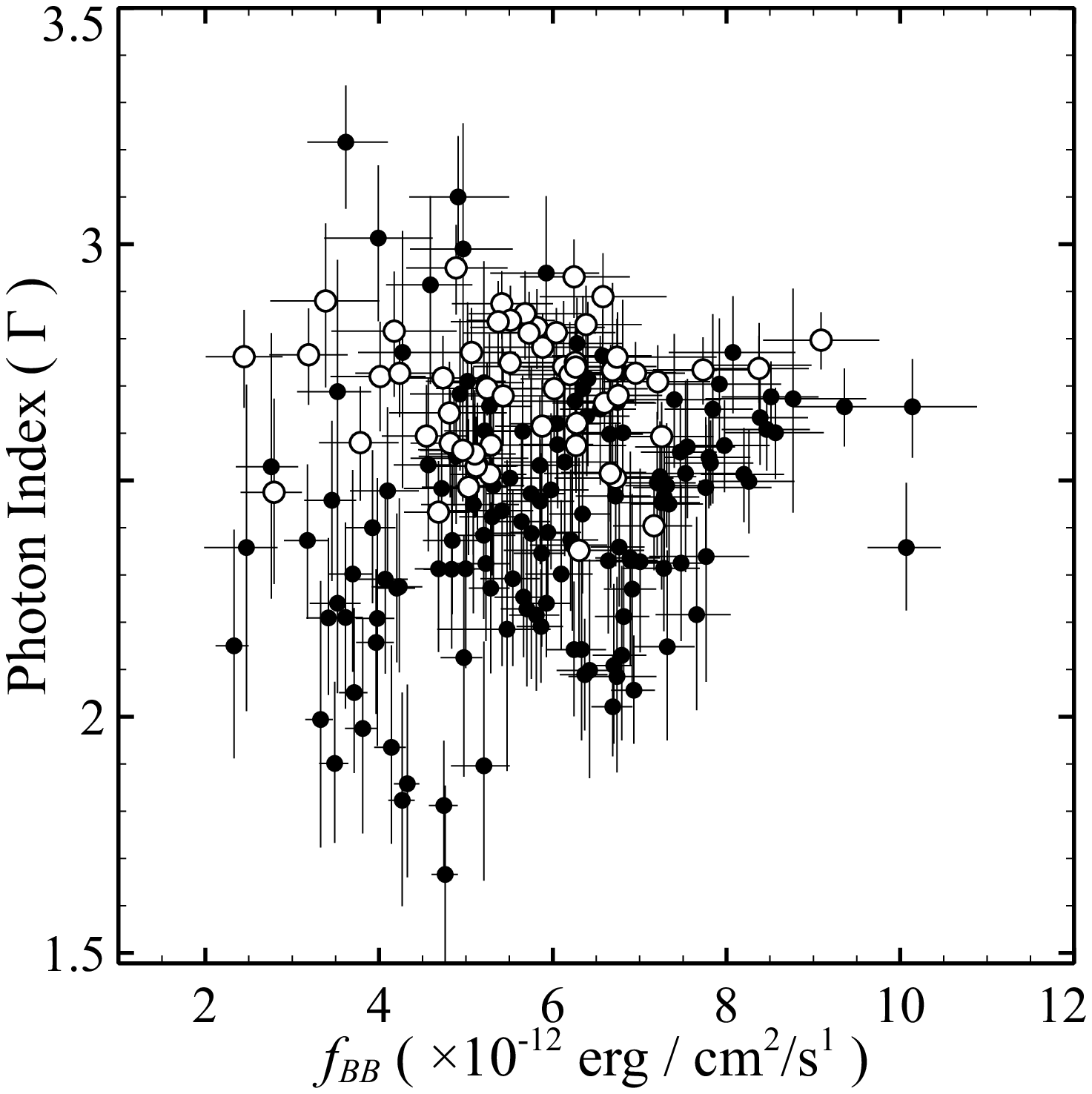}
  \includegraphics[scale=0.27]{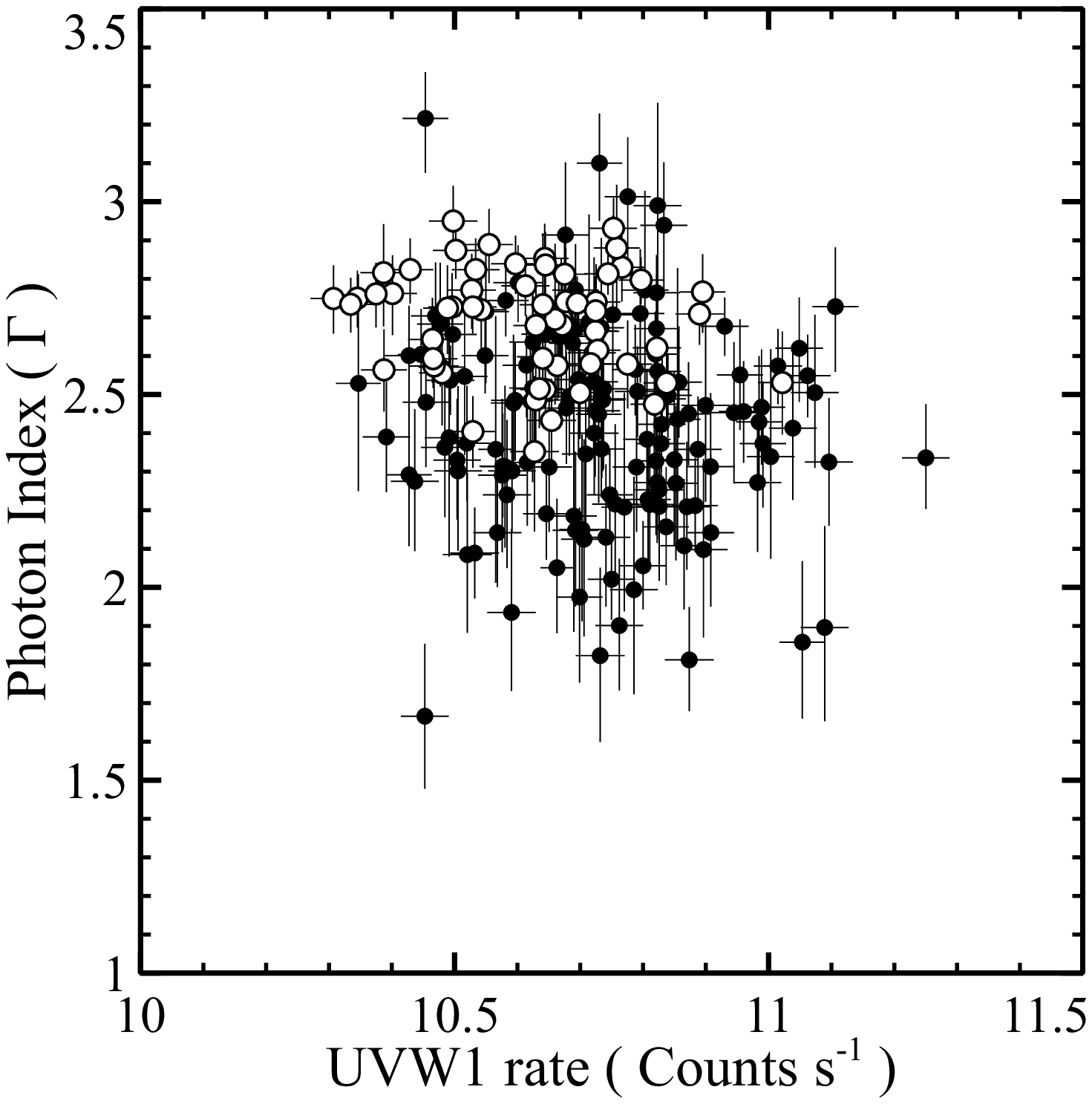}
  \caption{{\it Top panels}: The $f_{BB}$ vs $f_{PL}$ and the $\Gamma$ vs $f_{PL}$ relations. {\it Middle panels}: The kT$_{BB}$ vs $f_{PL}$ and the $f_{BB}$ vs the blackbody temperature (kT$_{BB}$), plots. {\it Bottom Panels:} The X--ray power
law photon index, $\Gamma$, plotted as a function of $f_{PL}$ (left), and UVW1 count rate (right). The filled(open) circles in all panels indicate the points which correspond to the left(right) group of points seen in the top--left panel (see text for details).}\label{spectral_variation}
\end{figure}

%%%%%%%%%%%%%%%%%%%%%%%%%%%%%%%
% Fig. 5
%%%%%%%%%%%%%%%%%%%%%%%%%%%%%%%%
\begin{figure}
  \centering
  \includegraphics[scale=0.27]{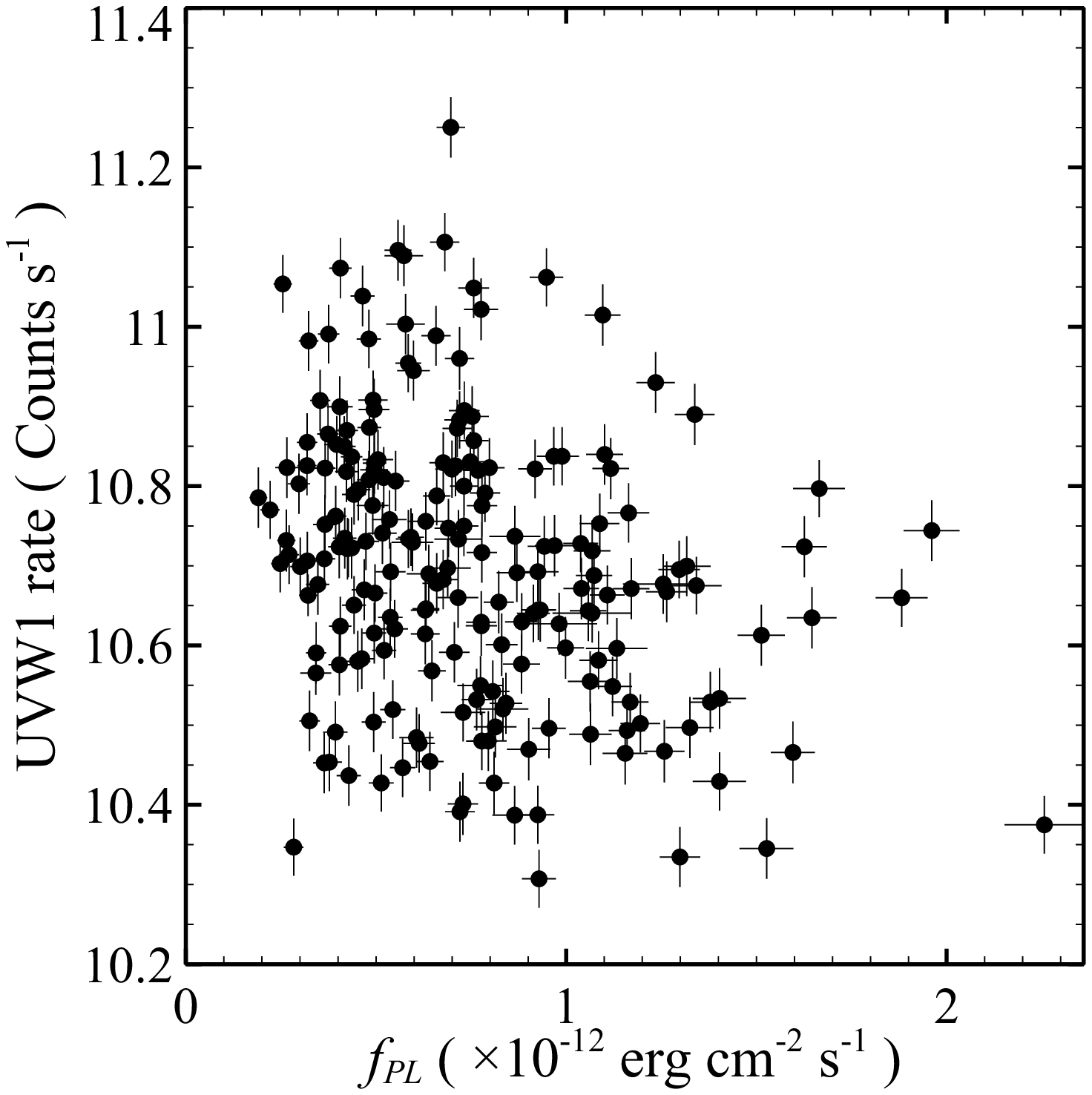}
  \includegraphics[scale=0.27]{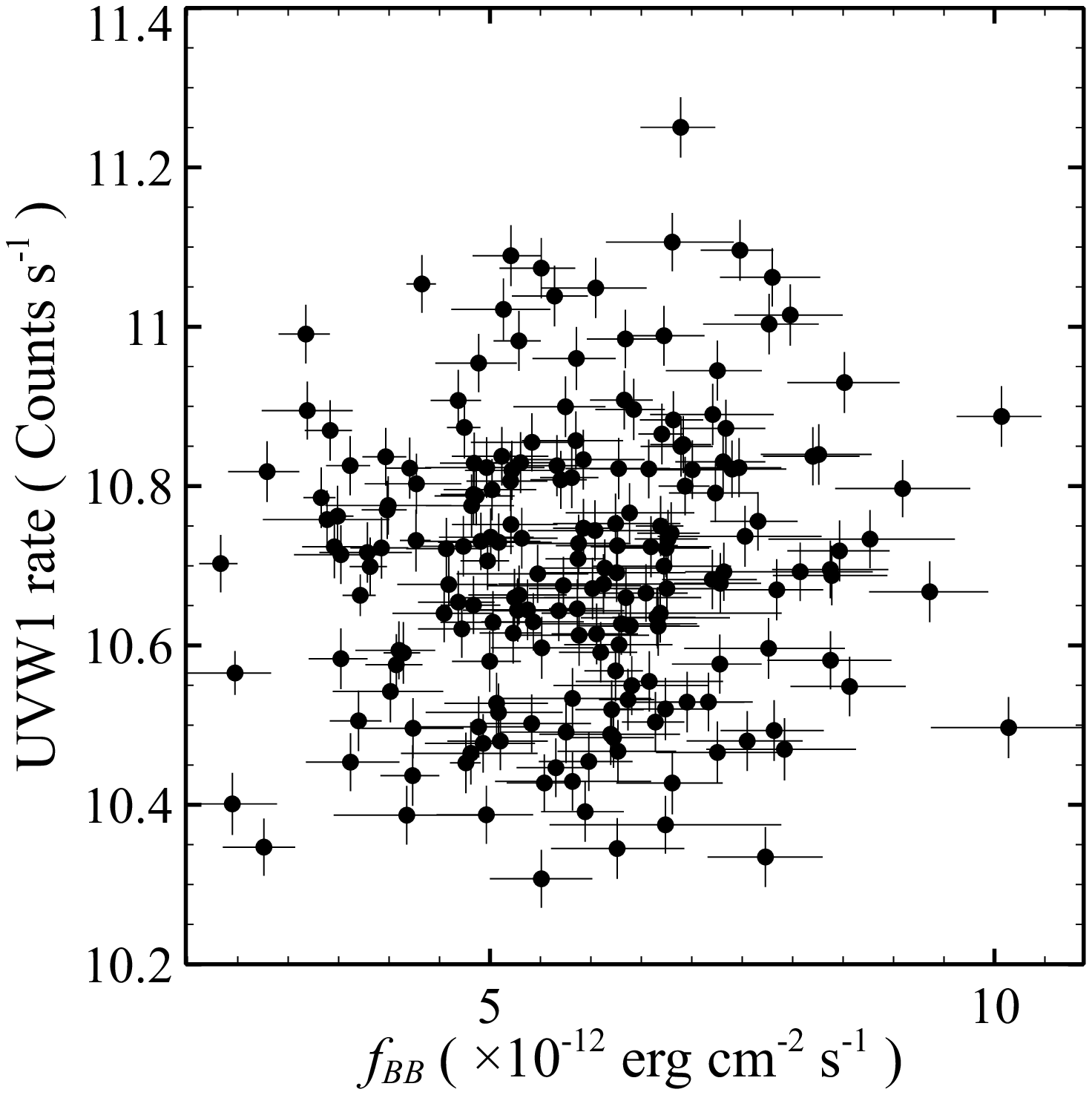}
  \caption{The UVW1 count rate vs $f_{PL}$ and $f_{BB}$ (left and right panels, respectively).}\label{uv_plbbflux}
\end{figure}

Figure~\ref{spectral_variation} shows the correlation between various best--fit spectral parameters. The errors correspond to the 1 $\sigma$ confidence level. The top--left panel in Fig.~\ref{spectral_variation} shows the correlation between the PL and the BB flux. This plot is analogous, but not identical, to the bottom right plot in Fig.~\ref{xmm_variability}. The SX count rate plotted in the later, is the sum of both the BB and the PL contribution in the soft band. In the present plot, we show the flux of the two components, in two separate bands. The PL and BB fluxes are well correlated. Interestingly, we observe two parallel branches of flux evolution, which are indicated by the filled and open circles (filled/open circles in all panels of this figure indicate the data in the two different branches of the $f_{BB}--f_{PL}$ plot, which we will refer to as the ``left" and the ``right--branch'' points, respectively). We defined the ``Hard Flux ratio”, HFR, as: HFR=$f_{PL}/f_{BB}$. This ratio is is indicative of the strength of the hard band PL with respect to the soft excess emission. The red line in the top right panel of Fig.~\ref{spectral_variation} indicates the HFR=0.15 line. All the right branch points are located on the right side of this line. In these observations, the hard band, PL flux is enhanced with respect to the soft excess flux. The appearance of two branches in this plot is not the result of the fact that out phenomenological model does not fit well some spectra. Most of the spectra where the best--fit $p_{null}$ is less than 0.01 fall on the left branch, where the majority of the points lay anyway.

The top--right panel in Fig.~\ref{spectral_variation} shows the PL spectral slope ($\Gamma$) vs $f_{PL}$. As is commonly observed in Seyfert 1 galaxies (e.g. \citealt{2009MNRAS.399.1597S}), the X–ray spectral slope is positively correlated with the PL flux with a correlation coefficient of 0.48 ($p_{null} = 3\times 10^{-13}$). The vertical green line indicates the $f_{pl}=8\times 10^{-13}$ \flux{} limit. Clearly, all the right branch points correspond to observations where the hard band flux is larger than 0.3 counts/s. There is a tight correlation between $f_{PL}$ and the HX count rate, and the $8\times 10^{-13}$ \flux{} PL flux limit corresponds to a HX rate of 0.3 counts/s. The horizontal line in the third panel of the Fig.~\ref{xmm_variability} indicates this limit.

In most cases (but not always), when the hard band count rate is larger than $\sim 0.3$ count/s and the source shows ``flare like'' peaks in its light curve, the increase of the $f_{PL}$ flux is stronger than the increase of the soft excess flux, so that $HFR>0.15$. At the same time, the PL spectral slope appears to be constant at $\Gamma\sim 2.7-2.8$, irrespective of the X—ray continuum flux.

The mid--left panel in Fig.~\ref{spectral_variation} shows the BB temperature (kT$_{BB}$) as a function of the PL flux. We observe a strong correlation between $f_{PL}$ and kT$_{BB}$. The correlation is strong for both the left and the right--branch points. We fitted the kT$_{BB} -- f_{PL}$ data with a straight line using the ordinary (Y/X) least square regression method, following \citep{1990ApJ...364..104I}. The resulting best--fit line is: kT$_{BB}=0.11(\pm 0.01)$~keV$ + 1.33(\pm0.11)\times 10^{10} f_{PL}$. The best--fit line intercept indicates the presence of a constant blackbody component on the probed timescales, which could be due to the disc thermal emission itself. 

The mid--right panel in Fig.~\ref{spectral_variation} shows that the BB flux broadly increases with increasing temperature. We estimate a correlation coefficient of $r=0.22$, which indicates a rather weak correlation between $f_{BB}$ and kT$_{BB}$ (although weak, the correlation is significant, as the null hypothesis probability, $p_{null}$, is $2\times 10^{-3}$). The fact that $f_{BB}$ broadly increases with increasing disc temperature is expected, since the BB flux should depend on the temperature. The fact that the correlation is not strong is probably due to the fact that the left-- and right--branch points follow two separate tracks in this plot, and the normalization of the right--branch $f_{BB}--$kT$_{BB}$ track (open circles) is smaller than the normalization of the left--branch track. Indeed, when we consider the right and left--branch points separately, the correlation coefficient increases to 0.46 and 0.29 ($p_{null}=2\times 10^{-4}$ and $4\times 10^{-4}$), respectively. 

The bottom panels in Fig.~\ref{spectral_variation} show the correlation plots between the best--fit PL $\Gamma$ vs $f_{BB}$, and the UVW1 count rate (left and right panels, respectively). The spectral slope variations do not correlate with the BB flux, when we consider all the points together ($r=0.15, p_{null}=0.04$). However, a significant (albeit weak), positive correlation is detected when we consider the left branch points only ($r=0.24, p_{null}=4\times10^{-3}$). As for the $\Gamma$ vs UVW1 correlation (right bottom panel in Fig.\,\ref{spectral_variation}, we detect a weak {\it anti--}correlation ($r=-0.23$) when we consider all data, which is significant ($p_{null}=1\times 10^{-3}$). The anti--correlation does not appear as significant, when we consider either the left or the right--branch points.

Finally, in Fig.\,\ref{uv_plbbflux} we plot the UVW1 count rate versus the best--fit PL and BB flux (left and right panels, respectively). The UVW1 vs $f_{PL}$ plot is almost identical to the UVW1 vs HX plot shown in Fig.\,\ref{xmm_variability}. This is not surprising, as $f_{PL}$ should be directly representative of the HX count rate. On the other hand, $f_{BB}$ accounts for just a portion of the SX count rate. We detect an $anti-$correlation between UVW1 count rate and $f_{PL}$ ($r=-0.27, p_{null}=1\times10^{-4}$). We reach a similar result when we consider the UVW1 vs HX plot as well: $r=-0.23, p_{null}=8\times 10^{-4}$. The UVW1--$f_{PL}$ anti--correlation can explain the $\Gamma-$UVW1 anti--correlation as well, since $\Gamma$ and $f_{PL}$ are positively correlated. On the other hand, we do not observed any significant correlation (positive or negative) between the UVW1 count rate and $f_{BB} (r=0.08, p_{null}=0.26$). Due to the fact that the BB emission contributes mainly in the SX band, the UVW1--SX count rate (anti)relation, is not as significant as the UVW1--HX anti--correlation ($r=-0.16, p_{null}=0.02)$. 

%====================================
\section{Discrete Cross--correlation Analysis}	
%====================================
%====================================
\subsection{The long term UV/X--ray cross--correlation}
%====================================

%%%%%%%%%%%%%%%%%%%%%%%%%%%%%%%
% Fig. 6
%%%%%%%%%%%%%%%%%%%%%%%%%%%%%%%%
\begin{figure*}
  \includegraphics[scale=0.45]{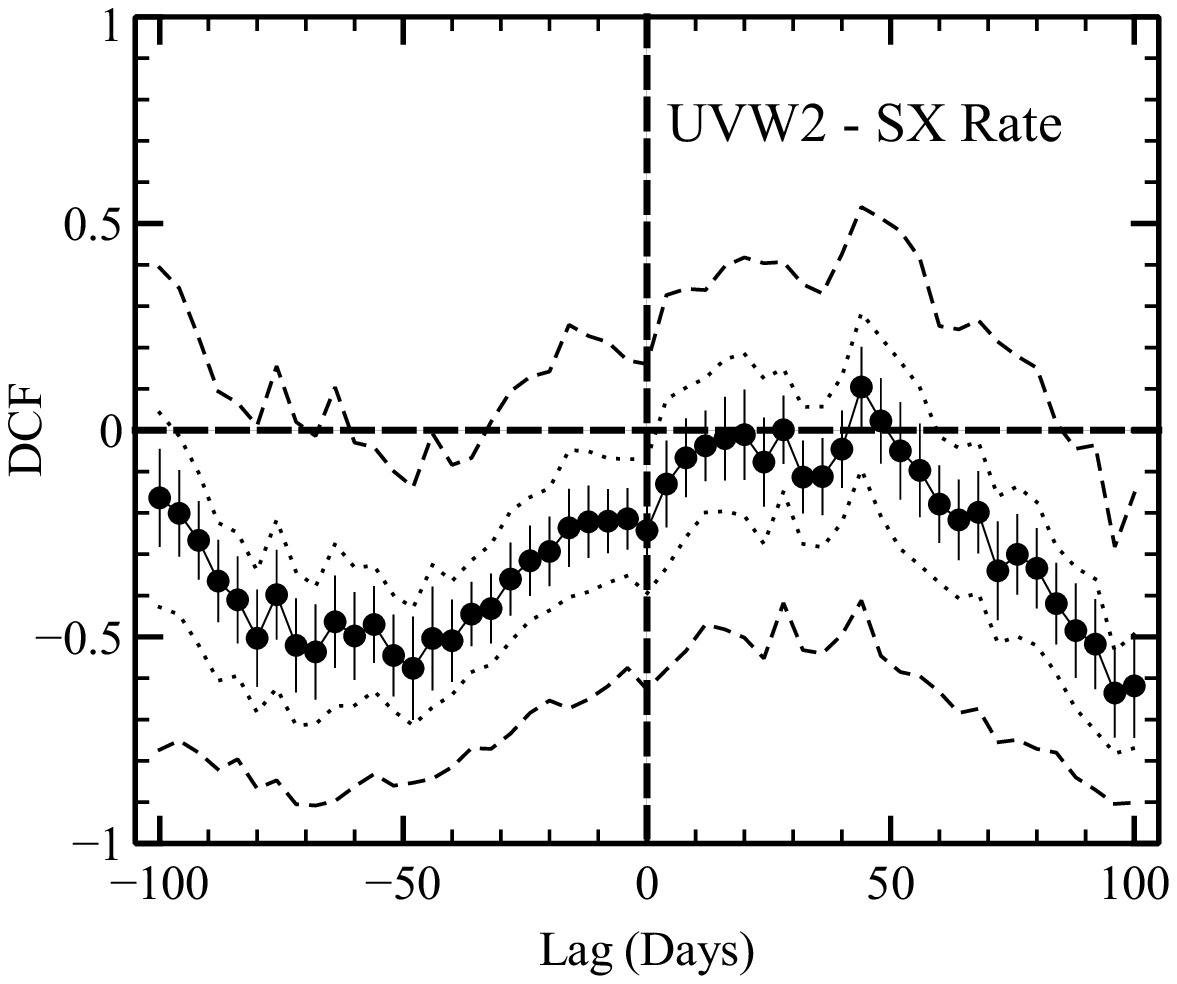}
  \includegraphics[scale=0.45]{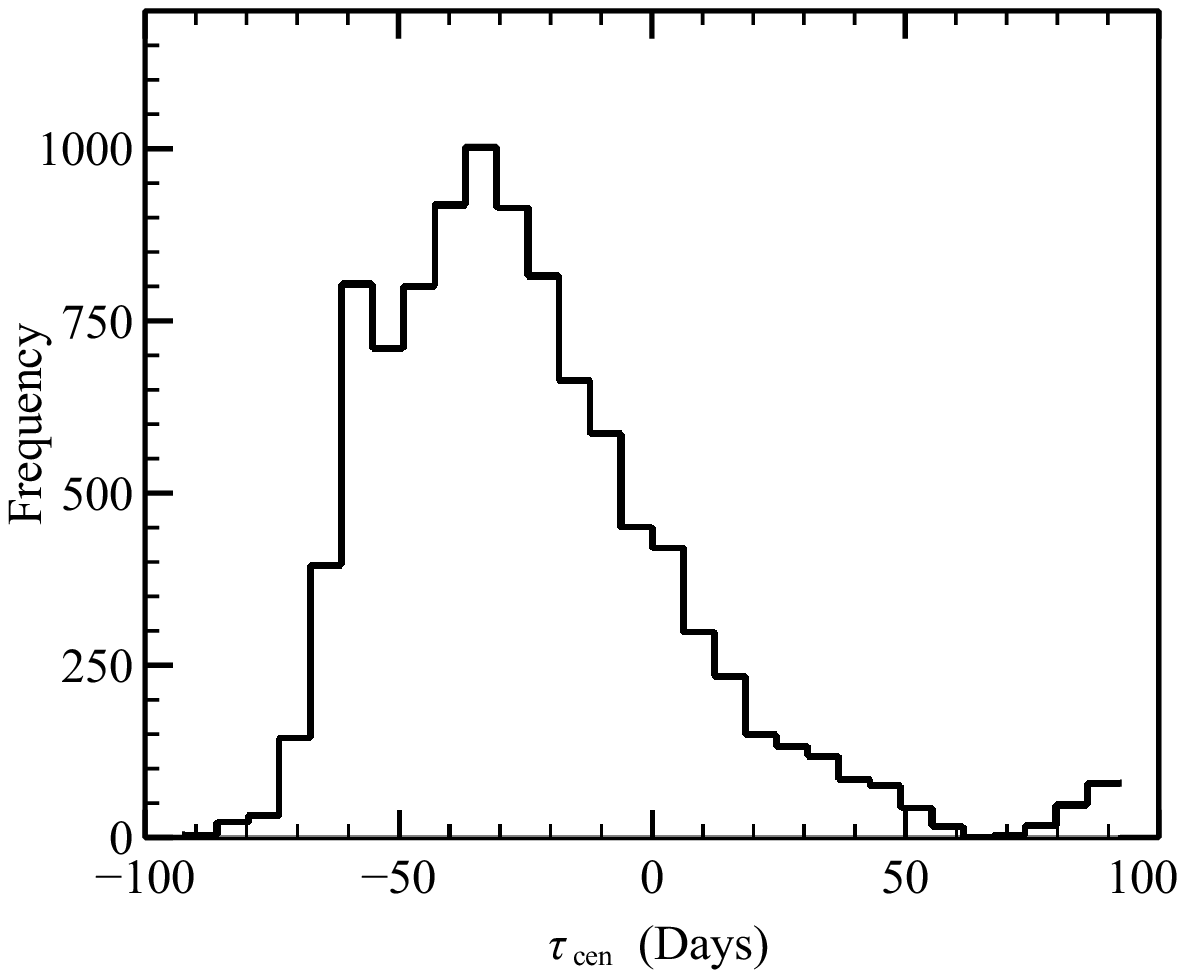}
  \includegraphics[scale=0.45]{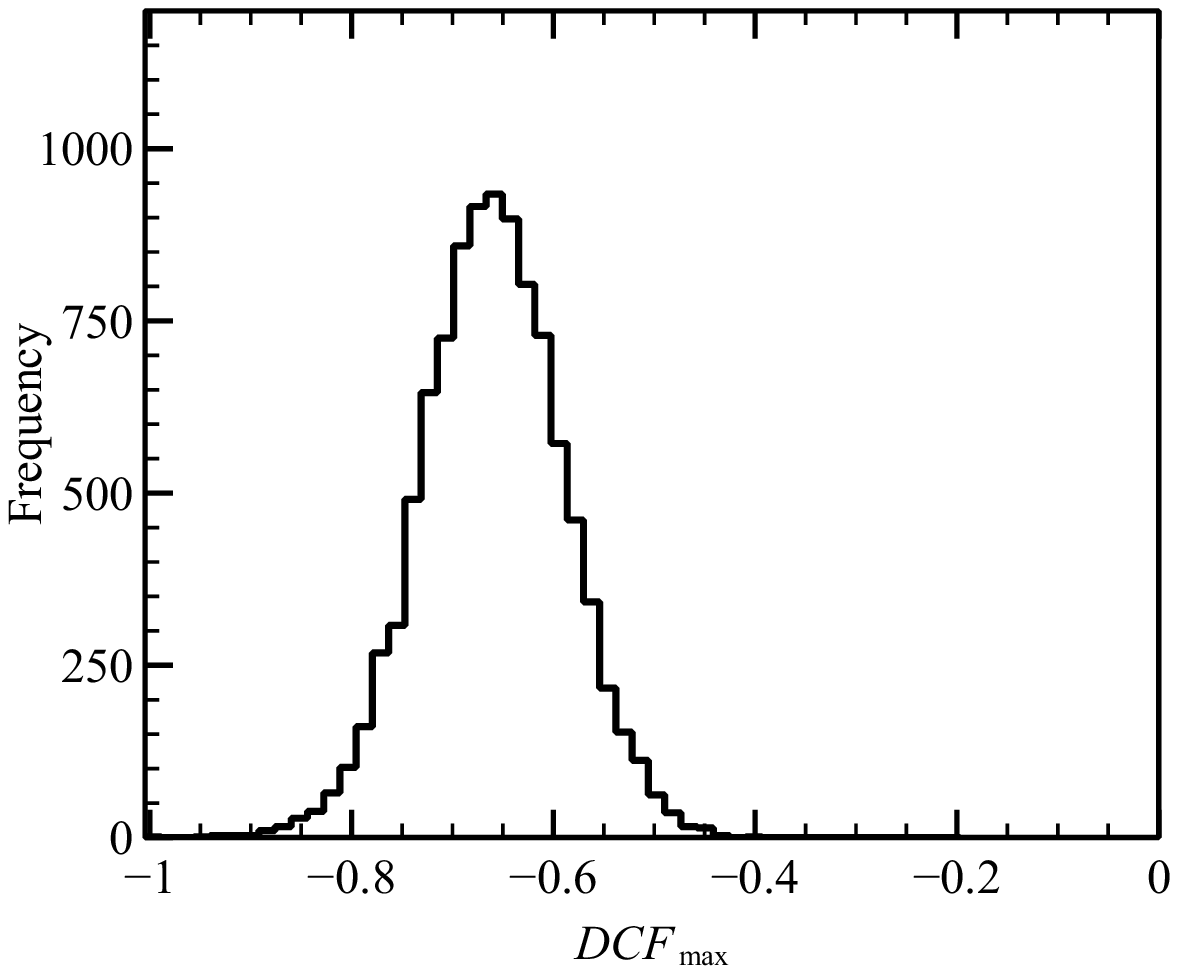}
  \caption{\emph{Left:} The DCF between \swift{} UVW2 and SX light curve using a lag size of 4 days. Lags are derived relative to the SX light curve, and positive lag means that SX lead UV variations. The dotted and dashed lines represents 68\% and 99.7\% confidence intervals obtained from 10000 simulations (see text for details). The centre and right panels show the probability distribution of $\tau_{cent}$ and DCF$_{max}$, respectively.}\label{dcf_swift} 
\end{figure*}

We computed cross--correlation functions using the Discrete Correlation Function (DCF) \citep{1988ApJ...333..646E}, as implemented in {\it Python} (i.e. {\it pydcf}). First, we considered the one year long \swift{} data to estimate the cross--correlation between the UV and the X--ray light curves on timescales of days/weeks. The \swift{} DCFs were calculated using a lag bin size of 4 days. The left~panel of Fig~\ref{dcf_swift} shows the DCF calculated using the \swift{} UVW2 and SX light curves (the UVW2/HX DCF is identical). The lags are such that a positive lag implies that the UV lag the X--ray variations. 

 We do not detect a positive correlation between the UV and X--ray bands, at almost any lag. Instead, we observe a moderate anti--correlation (DCF$_{\rm max}\sim -0.5$), with the UV leading the X--rays by $\sim 50$ days. We calculated the centroid of the DCF, $\tau_{\rm cent}$, as the mean of all the DCF points which are $0.8\times$DCF$_{\rm max}$, and we accepted it as our estimate of the time lag between two light curves. We also computed the average DCF$_{\rm max}$ as the mean of the DCF values of the same points. We found that $\tau_{\rm cent,UV/SX}=-56$ days and DCF$_{{\rm max, UV/SX}}=-0.54$ (similar results hold for the UV/HX DCF). A second negative DCF peak is also observed at +100 days, however this time lag is just three times smaller than the total duration of the light curves, hence its significance is rather uncertain. 
 
In order to estimate the significance of the detected anti--correlation we followed the Monte Carlo methods proposed by \cite{1998PASP..110..660P}. We created 10000 pairs of synthetic light curves, and we calculated the discrete correlation of each pair (DCF$_{simul}$). The dotted and dashed lines in the left panel of Fig~\ref{dcf_swift} indicate the 68\% and 99.7\% range of DCF$_{simul}$, in each lag. The negative DCF at lag $\sim -55$ days appears to be significant at the 3$\sigma$ level. We computed $\tau_{cent,simul}$ and DCF$_{max,simul}$ of all DCF$_{simul}$, exactly as we did with the observed light curves. We used these values to construct the respective sample distribution functions, which we assume are representative of the intrinsic distribution of $\tau_{\rm cent}$ and DCF$_{\rm max}$. The $\tau_{cent,simul}$ and DCF$_{max,simul}$ distributions are also plotted in Fig.~\ref{dcf_swift} (middle and right panels, respectively). $\tau_{cent,sim ul}$ distribution is very broad, while the DCF$_{max,simul}$ distribution is rather narrow. Using these distributions, we estimated the 95\% confidence limits on the observed centroid time lag and max DCF: $\tau_{cent,UV/SX}=-56.0^{+55}_{-38}$, DCF$_{\rm max,UV/SX}=-0.54^{+0.08}_{-0.10}$. We conclude that, on time scales longer than a few days, the \swift{} data imply an anti--correlation between the UV and the X--ray light curves in \hhh, but we cannot determine the delay accurately.

%====================================
\subsection{The short term UV/X--ray cross--correlation}\label{shortDCF}
%====================================

%%%%%%%%%%%%%%%%%%%%%%%%%%%%%%
% Fig. 7
\begin{figure}
%%%%%%%%%%%%%%%%%%%%%%%%%%%%%%%
  \centering
  \includegraphics[scale=0.5]{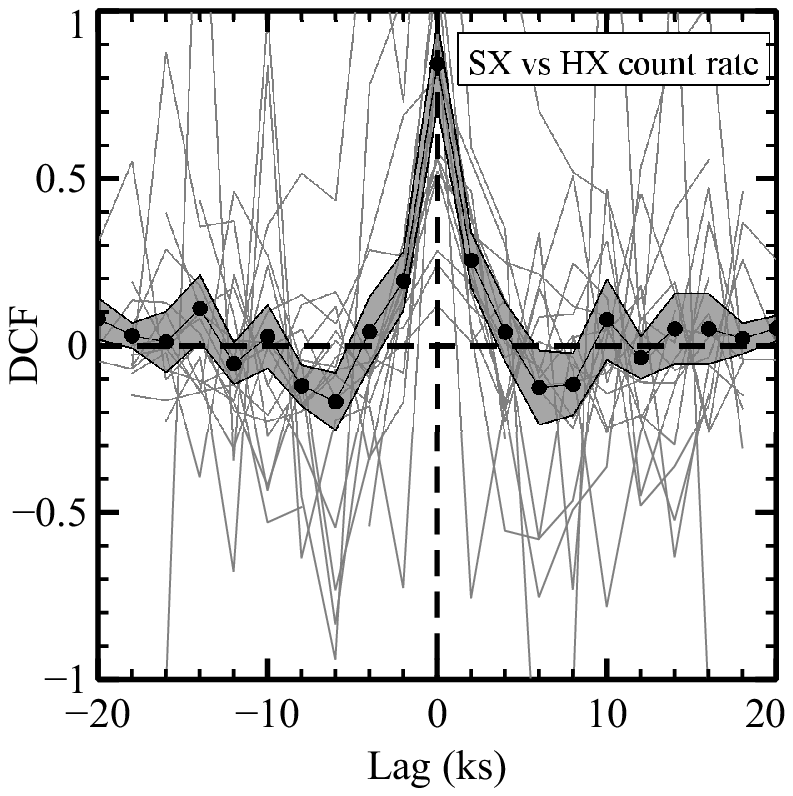}
  \includegraphics[scale=0.5]{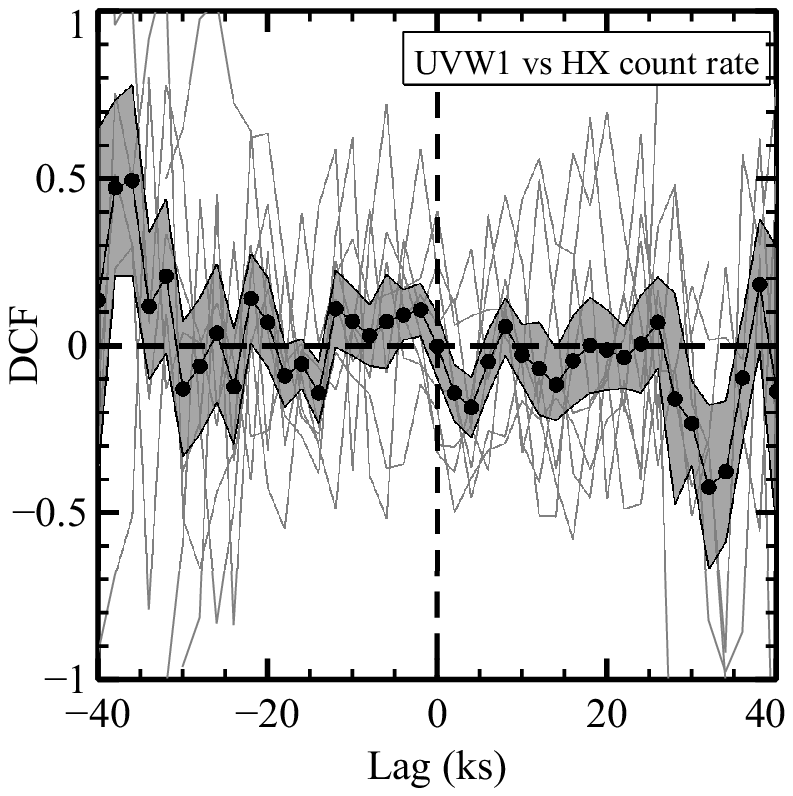}
  \caption{ Mean DCF between the HX and SX light curves (left panel), and between the UV count rate and the HX (right panel), using the \xmm{} data. The shaded area indicates the 1$\sigma$ error on the mean DCF. Lags are derived relative to the first band mentioned in each panel.}\label{dcf_xmm}
\end{figure}

We used the \xmm{} data to compute the DCFs between the HX and SX light curves, and between the UVW1 and the SX light curves. These DCFs are appropriate to study correlation on time scales of a few thousand seconds. We used the X--ray light curves binned like the OM light curves (see \S\,\ref{xmm_analysis}). To compute the X--ray DCF we chopped the SX and HX light curves in segments which were 20~ks long. We estimated the DCF for each one of them (using a lag bin size of 2~ks), and we computed the mean DCF at each lag. By chopping the light curve in short segments, and by subtracting each segment’s mean from the points in each segment when we estimate the DCF, we eliminate (to some extent) the variations on time scales longer than $\sim 20$~ks. This is a simple pre--whitening method, in order to estimate the time lags on time scales roughly equivalent, or smaller, than the size of each light curve segment. In addition, by suppressing the long term variations, the DCF of each segment is, to the extent, independent from the DCF of other segments (perhaps, for this reason, the sample DCFs of the individual segments look so different -- see Fig.\,\ref{dcf_xmm}). As a result, the mean DCF at each lag, and its error, could provide a more reliable estimate of the cross correlation of the short term variations, and its uncertainty. The mean X--ray DCF is shown in the left panel of Fig.\,\ref{dcf_xmm}.

In both panels, the grey lines indicate the individual DCFs, the filled circles show the mean DCF, and the shaded area is representative of the error of the average DCF. In both plots, a positive lag would indicate that the first mentioned band in the plot's label is leading the second band. The resulting DCFs conform the results we report in \S 2.3 We observe a strong, positive correlation at zero lag in the case of the HX vs SX correlation. The average UV/X--ray correlation is consistent with zero at all lags from $-40$ to +40~ks. Negative DCF peaks (which are not significant though) appear at lag$\sim0$ and $\sim +35$~ks. This is the right size for the delays, but it is the UV variations that are leading, and the negative DCF suggests an {\it anti--}correlation. 

%====================================
\section{Summary \& Discussion}	
%====================================
We investigated the UV/X--ray connection in \hhh{} on long and short time scales, using a year--long \swift{} monitoring observations during 2010--2011, and four long \xmm{} observations performed in 2008. We studied correlations between the UV flux and the count rate in the soft and hard X--ray bands. We used the \xmm{} data to perform time--resolved spectroscopy on short time scales, of the order of $\sim~1$--$2$~ks. We parametrized the X--ray continuum and the soft--excess with a simple PL and BB model, respectively. On short time scales, these models fully account for the 2--10~keV and the soft--excess flux, respectively. Using the results from the TRS study we investigated the connection between the two X--ray spectral components and with the UV.
We summarize our main results below, and we discuss their implications in the following sections. 

\begin{itemize}
\item We found significant UV and X--ray variations, on both long and short time scales. The variability amplitude increases from the UV, to the soft, and then to the hard X--ray band. In each band, the longer {\it Swift} light curves have a variability amplitude which is larger than the (shorter) {\it XMM-newton} light curves. This is typical behaviour for Seyferts.

\item We detect an anti--correlation between the UV and the X--ray flux variations, on both long and short timescales, although the long term anti--correlation is tentative (see below). The UV emission does not correlate with the BB flux (i.e. with soft excess). 

\item The BB and PL flux are positively correlated, but the correlation is complicated, as we observe two branches in the $f_{BB}$ vs $f_{PL}$ plot. 

\item The power law spectral slope steepens with increasing PL flux. The spectra slope does not correlate with the UVW1 count rate, while we observe a weak (but significant) positive correlation with $f_{BB}$.  

\item We observe a strong positive correlation between the black--body temperature and the power--law flux. 

\item The BB flux correlates positively with kT$_{BB}$, although the correlation is rather weak, with lots of scatter in the $f_{BB}--$kT$_{BB}$ plot.

\end{itemize}

%====================================
\subsection{The UV/X--ray connection}
%==================================== 

As we mentioned in the Introduction, numerous past studies have revealed a (weak) positive correlation between the UV and X--ray variations. In most cases, the X--ray leads the UV variations, while the opposite has been observed in a couple of sources. Our results show that \hhh{} is unique among the AGN that have been studied so far. 

On long time--scales, we find the UV flux to be variable, and at the same time, to be {\it anti-}correlated with the X--ray variations, with the UV leading the X--rays by about $\sim 50$ days (the error on the estimated delay is large, and a zero delay cannot be rejected). We inspected visually the DCF of many pairs of simulated light curves and we saw that, in almost all cases, they are predominantly negative, with narrow, large amplitude peaks, which appear over a very broad range of time lags. This explains the broad distribution of $\tau_{\rm cent}$ but also the rather narrow distribution of DCF$_{\rm max}$ at values smaller than --0.5. Our results are consistent with \citealt{2015MNRAS.453.3455R}, who detected an anti--correlation between UV and X--rays, with the UV leading the X--rays by about $\sim 6$ days. Although the anti--correlation appears to be significant, results like e.g. periodicities (or delays, in our case) which are based on a small number of cycles in the observed light curves do not necessarily correspond to an intrinsic time scale in the case of variations which are due to a red--noise random process (as is the case with the AGN X--ray and optical/UV light curves; e.g. \citep{2006ASPC..360..101U}. Due to the red noise nature of the light curves variations may look like they are correlated (or anti--correlated) with some delay, just by chance. False correlations appear when the light curves are not far longer than the observed ``signal". In our case, the delay of $\sim 55$ days is just 7 times shorter than the total length of the {\it Swift} light curves, perhaps not large enough to be certain it is real.

What is though clear is that the long term variations of the source are {\it not} positively correlated. 
This result is certainly not consistent with the hypothesis of X--ray reprocessing being the main driver of the observed UV variations on long time scales. This is puzzling, given that the evidence for X--ray disc reflection is strong in this object (see section~\ref{subsection1}). One possible explanation for the lack of positive correlation between UV and X--rays is that the X--ray source is located very close to the central BH. In this case, the X--rays should be strongly ``beamed'' (bent) towards the inner disc, and will not be able to illuminate the outer part of the disc, where the UV emission is produced. At the same time, the UV/X--ray anti--correlation (or no positive correlation) is also not consistent with the hypothesis that the variations are produced in the outer part of the accretion flow (due to random fluctuations in the accretion rate) and then propagate inwards, affecting first the UV emitting region and then the X--ray light curves. 

The possibility of a UV/X--rays anti--correlation is enhanced by the fact that we also find significant evidence for a similar anti--correlation in the XMM--data as well. The UVW1 vs HX DCF does not show a significant negative peak at time scales shorter than half a day or so, but the correlation coefficient between the UVW1 and HX count rate (or $f_{PL}$) strongly suggests an anti--correlation, even at time scales as short as a few thousand second. We cannot provide any reasonable explanation of this anti--correlation. Further monitoring of the source (in both UV and X--rays), will be helpful to understand this puzzling result. 

In any case, the anti--correlation is relatively weak: $\sim -0.3$ on short time scales, and $\sim -0.5$ on longer time scales. This result suggests that the main variability mechanisms in the innermost and in the outer disc regions, which regulate most of the observed X--ray and UV variations, are independent of each other. Whatever is the physical cause for the X--ray variations variations, it does not appear to affect the UV emitting region. Therefore, the question as to which is the physical mechanism responsible for the UV variations in \hhh{} is still open.

%====================================
\subsection{Absorption induced UV variability} 
%====================================

One possibility could be clouds (like the broad line region (BLR) clouds, for example) moving across our line of sight. In fact, such BLR clouds have been invoked to explain the eclipsing X--ray variability of some AGN such as NGC~1365 (\citealt{2005ApJ...623L..93R}; \citealt{2007ASPC..373..447M}). If these clouds are in the form of a flattened ring or torus like geometry, they can eclipse the central X--ray source if the inclination is large (e.g., the case of the intermediate Seyfert NGC~1365). In case of low inclination, the clouds may obscure the disc at large radii (where the UV continuum emission originates), while leaving the central X--ray source unattenuated. Such a geometry may at work in \hhh, where there is no evidence of intrinsic X--ray absorption. If the UV variability is really caused by the moving BLR clouds, we expect to see optical/UV color variations correlated with the UV flux. Unfortunately, the \xmm{} and \swift{} observations were performed in a single filter and the possibility of colour variations could not be examined. Simultaneous observations in two optical/UV bands and in the X--ray band, such as those possible with the multi--wavelength mission {\it AstroSat} for example, will be able to probe possible absorption/extinction of the UV emission by the BLR clouds.

%====================================
\subsection{Soft excess as X--ray reprocessed thermal emission}\label{subsection1}
%====================================
The detection of broad iron K and L lines, and the $30$~s time delay between the primary continuum and the soft band X--ray emission containing the iron L line have clearly demonstrated the presence of relativistically blurred reflection arising from the innermost disc in \hhh{} \citep{2002MNRAS.329L...1B,2009Natur.459..540F}. Since not all X--rays are reflected, the strong illumination should result in reprocessed disc emission, which should be closely linked with the primary X--ray emission. Some of our results are fully consistent with X--ray reprocessing taking place in \hhh. 

For example, we find that the soft excess in this object is well reproduced by a BB component 
with a temperature of $\sim 0.11-0.14$~keV. As we discussed in \S \ref{TRSresults}, some of the BB flux should originate from the intrinsic disc thermal emission itself. Disc emission at such high energies is expected in the case of highly accreting sources with a maximally spinning central BH (see for example Fig.\,1 in \citep{2012MNRAS.426..656S}: the SED of a disc around a maximally spinning, $10^7$ M$_{\odot}$ BH, peaks at energies even higher than 0.1~keV, when the accretion rate is 0.3 of the Eddington limit). 

The BB temperature is variable and is positively correlated with the PL continuum flux (middle left panel of Fig.\,\ref{spectral_variation}), with no delay\footnote{Although not shown in the paper, we estimated the DCF between kT$_{BB}$ and $f_{PL}$ using the results from the TRS analysis, and 20~ks long segments (as in the DCFs discussed in \S \ref{shortDCF}. The DCF shows a strong peak at zero lag, suggesting that the disc temperature reacts with no delay to the X--ray continuum flux variations, as it is expected in the case of X--ray reprocessing and the X--ray source located close to the inner disc}. This is what we would expect if the soft excess flux is due to X--reprocessing on the disc. In this case, the part of the X--ray flux which is not reflected off the disc is absorbed and can increase the disc temperature. As a result, the disc could emit in the soft X--ray band. In this case, it is natural to expect a positive correlation between the PL flux and kT$_{BB}$. 

The BB flux should depend on kT$_{BB}$ and the area of the emitting region. Since the BB temperature responds to the primary flux, the BB flux should also be variable and should also correlate, positively, with the X--ray primary flux. This is indeed what we observe (left top panel in Fig.\,\ref{spectral_variation}). However, the correlation is rather complex, showing a possible bi--modal behaviour. Most of the times, when the primary X—ray continuum flux exceeds $\sim 8\times 10^{-13}$ \flux{}, the PL over the soft excess flux ratio increases. These high PL fluxes are associated with short, flare--like events.

But, we cannot be certain whether the source indeed operates in two distinctive modes, as the middle right panel in Fig.\,\ref{spectral_variation} suggests a slightly different picture. The bi--modal behaviour is not very clear in the $f_{BB}-kT_{BB}$ plane, as at any given $kT_{BB}$ (and hence $f_{PL}$) we observe a range of soft--excess flux values. The less than perfect, noisy $f_{BB}-kT_{BB}$ correlation can be explained if the soft--excess emitting surface area is also variable. The right--branch points could correspond to a smaller emitting area when compared to the left branch points with the same $kT_{BB}$ (and hence $f_{PL}$). To investigate this possibility further, we fitted the left and right branch data in the middle, right panel in Fig.\,\ref{spectral_variation} with a straight line, using the OLS(Y$|$X) method of \cite{1990ApJ...364..104I}. We found that the best--fit lines have the same slope (i.e. they are parallel), while the ratio of the left branch over the right branch best--fit normalisation is $\sim 1.1$. Since the BB flux should be proportional to $R^2T_{BB}^4$, the ratio of normalizations should be proportional to the square of the average ratio of the emitting region size in the left and right branch ``states''. Our results indicate that, on average, an increase in the size of the emitting region by a factor of $\sim 5$\% can explain the mean difference between the left and right branch points (for the same BB temperature). This situation points to a ``dynamic'' corona, with a variable geometry. If, the height and/or the size of the X--ray corona are randomly variable (irrespective of the flux), the size of the disc reprocessing region will also vary accordingly, explaining the noisy, broad correlation between $f_{BB}$ and $kT_{BB}$.

%We cannot be certain whether the source indeed operates in two different modes, as the middle right panel in Fig.\,\ref{spectral_variation} suggests a slightly different picture. The less than perfect, noisy $f_{BB}-$kT$_{BB}$ correlation can be explained if the soft-excess emitting surface area is also variable. The right-branch points could correspond to a smaller emitting area, which, for the same kT$_{BB}$ (and hence $f_{PL}$), should result in smaller BB flux. However, the bi-modal behaviour is not very clear in the $f_{BB}-$kT$_{BB}$ plane. The scatter of the correlation in this plot suggests that, for a given primary flux, the size of the disc which reprocesses the X--ray primary flux does not flip from a maximum to a minimum value (in a bi-modal way). Instead, it may vary in a rather continuous way between a max-min value, resulting in a range of soft-excess flux values, at a given $f_{PL}$ (and hence kT$_{BB}$).This situation points to a ```dynamic'' corona, with a variable geometry. If, the height and/or the size of the X--ray corona are randomly variable (irrespective of the flux), the size of the disc reprocessing region will also vary accordingly, explaining the noisy, broad correlation between $f_{BB}-$kT$_{BB}$, and $f_{PL}-f_{BB}$.
 
%====================================
\subsection{Spectral variability and the seed photons for Comptonization}
%====================================
The photon index of the X--ray power law is strongly correlated with the
power--law flux, without any measurable delay\footnote{Although not shown in the paper, we estimated the DCF between $f_{PL}$ and $\Gamma$ using the results from the TRS analysis, and 20~ks long segments (as in the DCFs discussed in \S \ref{shortDCF} The DCF shows a strong peak at zero lag}. The power law is found to steepen at higher flux. This
type of spectral behaviour is common in Seyfert 1 galaxies studied by 
\cite{2009MNRAS.399.1597S} and other authors e.g. 3C~120 (\citealt{2001ApJ...551..186Z}),
 NGC~3227 (\citealt{2009ApJ...691..922M}), Ton~S180 (\citealt{2002ApJ...564..162R}), 
NGC~7469 (\citealt{1998ApJ...505..594N}). 

This behaviour can originate by variations of a complex absorber (as e.g., in the case of NGC~1365), but lack of absorption features in the X--ray spectrum of \hhh{} rules out this possibility. Another possibility is that the spectral variability is the result of variable power law normalization and a relatively steady blurred reflection. However, our spectral slope measurements have been made by taking into account the independently variable soft X--ray excess component, so this is not a viable explanation.

This steeper when brighter behaviour can also be explained in the case of thermal Comptonization in a hot plasma irradiated by variable seed photons (e.g., \citealt{2001ApJ...551..186Z}). An increase in the input photons flux can cool the X--ray corona, thus producing a steepening of the spectral slope $\Gamma$. We therefore expect a steepening of the spectrum with increasing soft input flux. However, as we discussed in \S \ref{TRSresults}, we detect an $anti-$correlation between UVW1 and $\Gamma$, on short time scales. On longer time scales, we do not observe any correlation between UV and spectral slope variations (when we estimated the correlation coefficient between UVW2 and HR, we found $r=-0.28$, with $p_{null}=0.01$). We therefore conclude that neither the UVW2 nor the UVW1 photons are the majority of the input photons to the X--ray corona. 

On the other hand, we argued above that strong X--ray illumination of the inner disc takes place in \hhh, and is responsible for the (variable) soft excess in this source. In this case, we would expect the soft excess emission to contribute a significant part of (if not the whole of) the soft input photons to the corona. We observe a significant positive correlation between $f_{BB}$ and $\Gamma$ (see \S \ref{spectral_variation}). This is the kind of correlation we would expect if the BB (i.e. the soft excess) photons were the input photons to the hot corona: as their flux increases, it may cool the corona and therefore result in steeper spectral slopes. However, the correlation is weak though, which indicates that, although a small part of the observed $\Gamma$ variations may be caused by the variable soft excess (which in turn is produced by the X--ray primary reprocessed emission), the majority of the observed spectral variations should be intrinsic to the dynamic, hot corona.

\section{Acknowledgements}
PKP acknowledges financial support from CSIR, New Delhi, India.\\

\def\aj{AJ} \def\actaa{Acta Astron.}  \def\araa{ARA\&A} \def\apj{ApJ}
\def\apjl{ApJ} \def\apjs{ApJS} \def\ao{Appl.~Opt.}  \def\apss{Ap\&SS}
\def\aap{A\&A} \def\aapr{A\&A~Rev.}  \def\aaps{A\&AS} \def\azh{AZh}
\def\baas{BAAS} \def\bac{Bull. astr. Inst. Czechosl.}
\def\caa{Chinese Astron. Astrophys.}  \def\cjaa{Chinese
  J. Astron. Astrophys.}  \def\icarus{Icarus} \def\jcap{J. Cosmology
  Astropart. Phys.}  \def\jrasc{JRASC} \def\mnras{MNRAS}
\def\memras{MmRAS} \def\na{New A} \def\nar{New A Rev.}
\def\pasa{PASA} \def\pra{Phys.~Rev.~A} \def\prb{Phys.~Rev.~B}
\def\prc{Phys.~Rev.~C} \def\prd{Phys.~Rev.~D} \def\pre{Phys.~Rev.~E}
\def\prl{Phys.~Rev.~Lett.}  \def\pasp{PASP} \def\pasj{PASJ}
\def\qjras{QJRAS} \def\rmxaa{Rev. Mexicana Astron. Astrofis.}
\def\skytel{S\&T} \def\solphys{Sol.~Phys.}  \def\sovast{Soviet~Ast.}
\def\ssr{Space~Sci.~Rev.}  \def\zap{ZAp} \def\nat{Nature}
\def\iaucirc{IAU~Circ.}  \def\aplett{Astrophys.~Lett.}
\def\apspr{Astrophys.~Space~Phys.~Res.}
\def\bain{Bull.~Astron.~Inst.~Netherlands}
\def\fcp{Fund.~Cosmic~Phys.}  \def\gca{Geochim.~Cosmochim.~Acta}
\def\grl{Geophys.~Res.~Lett.}  \def\jcp{J.~Chem.~Phys.}
\def\jgr{J.~Geophys.~Res.}
\def\jqsrt{J.~Quant.~Spec.~Radiat.~Transf.}
\def\memsai{Mem.~Soc.~Astron.~Italiana} \def\nphysa{Nucl.~Phys.~A}
\def\physrep{Phys.~Rep.}  \def\physscr{Phys.~Scr}
\def\planss{Planet.~Space~Sci.}  \def\procspie{Proc.~SPIE}
\let\astap=\aap \let\apjlett=\apjl \let\apjsupp=\apjs \let\applopt=\ao

\bibliographystyle{mn} 
\bibliography{mn_17_1313_rev2}

\begin{thebibliography}{57}
\expandafter\ifx\csname natexlab\endcsname\relax\def\natexlab#1{#1}\fi

\bibitem[{{Ar{\'e}valo} {et~al.}(2005){Ar{\'e}valo}, {Papadakis}, {Kuhlbrodt},
  \& {Brinkmann}}]{2005A&A...430..435A}
{Ar{\'e}valo}, P., {Papadakis}, I., {Kuhlbrodt}, B., \& {Brinkmann}, W., 2005,
  \aap, 430, 435

\bibitem[{{Boller} {et~al.}(1996){Boller}, {Brandt}, \&
  {Fink}}]{1996A&A...305...53B}
{Boller}, T., {Brandt}, W.~N., \& {Fink}, H., 1996, \aap, 305, 53

\bibitem[{{Boller} {et~al.}(2002){Boller}, {Fabian}, {Sunyaev}, {Tr{\"u}mper},
  {Vaughan}, {Ballantyne}, {Brandt}, {Keil}, \&
  {Iwasawa}}]{2002MNRAS.329L...1B}
{Boller}, T., {Fabian}, A.~C., {Sunyaev}, R., {et~al.}, 2002, \mnras, 329, L1

\bibitem[{{Buisson} {et~al.}(2017){Buisson}, {Lohfink}, {Alston}, \&
  {Fabian}}]{2017MNRAS.464.3194B}
{Buisson}, D.~J.~K., {Lohfink}, A.~M., {Alston}, W.~N., \& {Fabian}, A.~C.,
  2017, \mnras, 464, 3194

\bibitem[{{Burrows} {et~al.}(2005){Burrows}, {Hill}, {Nousek}, {Kennea},
  {Wells}, {Osborne}, {Abbey}, {Beardmore}, {Mukerjee}, {Short}, {Chincarini},
  {Campana}, {Citterio}, {Moretti}, {Pagani}, {Tagliaferri}, {Giommi},
  {Capalbi}, {Tamburelli}, {Angelini}, {Cusumano}, {Br{\"a}uninger}, {Burkert},
  \& {Hartner}}]{2005SSRv..120..165B}
{Burrows}, D.~N., {Hill}, J.~E., {Nousek}, J.~A., {et~al.}, 2005, \ssr, 120,
  165

\bibitem[{{Dauser} {et~al.}(2012){Dauser}, {Svoboda}, {Schartel}, {Wilms},
  {Dov{\v c}iak}, {Ehle}, {Karas}, {Santos-Lle{\'o}}, \&
  {Marshall}}]{2012MNRAS.422.1914D}
{Dauser}, T., {Svoboda}, J., {Schartel}, N., {et~al.}, 2012, \mnras, 422, 1914

\bibitem[{{Dewangan} {et~al.}(2007){Dewangan}, {Griffiths}, {Dasgupta}, \&
  {Rao}}]{2007ApJ...671.1284D}
{Dewangan}, G.~C., {Griffiths}, R.~E., {Dasgupta}, S., \& {Rao}, A.~R., 2007,
  \apj, 671, 1284

\bibitem[{{Dickey} \& {Lockman}(1990)}]{1990ARA&A..28..215D}
{Dickey}, J.~M. \& {Lockman}, F.~J., 1990, \araa, 28, 215

\bibitem[{{Done} {et~al.}(2012){Done}, {Davis}, {Jin}, {Blaes}, \&
  {Ward}}]{2012MNRAS.420.1848D}
{Done}, C., {Davis}, S.~W., {Jin}, C., {Blaes}, O., \& {Ward}, M., 2012,
  \mnras, 420, 1848

\bibitem[{{Edelson} {et~al.}(2015){Edelson}, {Gelbord}, {Horne}, {McHardy},
  {Peterson}, {Ar{\'e}valo}, {Breeveld}, {De Rosa}, {Evans}, {Goad}, {Kriss},
  {Brandt}, {Gehrels}, {Grupe}, {Kennea}, {Kochanek}, {Nousek}, {Papadakis},
  {Siegel}, {Starkey}, {Uttley}, {Vaughan}, {Young}, {Barth}, {Bentz},
  {Brewer}, {Crenshaw}, {Dalla Bont{\`a}}, {De Lorenzo-C{\'a}ceres}, {Denney},
  {Dietrich}, {Ely}, {Fausnaugh}, {Grier}, {Hall}, {Kaastra}, {Kelly},
  {Korista}, {Lira}, {Mathur}, {Netzer}, {Pancoast}, {Pei}, {Pogge},
  {Schimoia}, {Treu}, {Vestergaard}, {Villforth}, {Yan}, \&
  {Zu}}]{2015ApJ...806..129E}
{Edelson}, R., {Gelbord}, J.~M., {Horne}, K., {et~al.}, 2015, \apj, 806, 129

\bibitem[{{Edelson} \& {Krolik}(1988)}]{1988ApJ...333..646E}
{Edelson}, R.~A. \& {Krolik}, J.~H., 1988, \apj, 333, 646

\bibitem[{{Epitropakis} {et~al.}(2016){Epitropakis}, {Papadakis}, {Dov{\v
  c}iak}, {Pech{\'a}{\v c}ek}, {Emmanoulopoulos}, {Karas}, \&
  {McHardy}}]{2016A&A...594A..71E}
{Epitropakis}, A., {Papadakis}, I.~E., {Dov{\v c}iak}, M., {Pech{\'a}{\v c}ek},
  T., {Emmanoulopoulos}, D., {Karas}, V., \& {McHardy}, I.~M., 2016, \aap, 594,
  A71

\bibitem[{{Fabian} {et~al.}(2002){Fabian}, {Ballantyne}, {Merloni}, {Vaughan},
  {Iwasawa}, \& {Boller}}]{2002MNRAS.331L..35F}
{Fabian}, A.~C., {Ballantyne}, D.~R., {Merloni}, A., {Vaughan}, S., {Iwasawa},
  K., \& {Boller}, T., 2002, \mnras, 331, L35

\bibitem[{{Fabian} {et~al.}(2009){Fabian}, {Zoghbi}, {Ross}, {Uttley}, {Gallo},
  {Brandt}, {Blustin}, {Boller}, {Caballero-Garcia}, {Larsson}, {Miller},
  {Miniutti}, {Ponti}, {Reis}, {Reynolds}, {Tanaka}, \&
  {Young}}]{2009Natur.459..540F}
{Fabian}, A.~C., {Zoghbi}, A., {Ross}, R.~R., {et~al.}, 2009, \nat, 459, 540

\bibitem[{{Fabian} {et~al.}(2012){Fabian}, {Zoghbi}, {Wilkins}, {Dwelly},
  {Uttley}, {Schartel}, {Miniutti}, {Gallo}, {Grupe}, {Komossa}, \&
  {Santos-Lle{\'o}}}]{2012MNRAS.419..116F}
{Fabian}, A.~C., {Zoghbi}, A., {Wilkins}, D., {et~al.}, 2012, \mnras, 419, 116

\bibitem[{{Gehrels} {et~al.}(2004){Gehrels}, {Chincarini}, {Giommi}, {Mason},
  {Nousek}, {Wells}, {White}, {Barthelmy}, {Burrows}, {Cominsky}, {Hurley},
  {Marshall}, {M{\'e}sz{\'a}ros}, {Roming}, {Angelini}, {Barbier}, {Belloni},
  {Campana}, {Caraveo}, {Chester}, {Citterio}, {Cline}, {Cropper}, {Cummings},
  {Dean}, {Feigelson}, {Fenimore}, {Frail}, {Fruchter}, {Garmire}, {Gendreau},
  {Ghisellini}, {Greiner}, {Hill}, {Hunsberger}, {Krimm}, {Kulkarni}, {Kumar},
  {Lebrun}, {Lloyd-Ronning}, {Markwardt}, {Mattson}, {Mushotzky}, {Norris},
  {Osborne}, {Paczynski}, {Palmer}, {Park}, {Parsons}, {Paul}, {Rees},
  {Reynolds}, {Rhoads}, {Sasseen}, {Schaefer}, {Short}, {Smale}, {Smith},
  {Stella}, {Tagliaferri}, {Takahashi}, {Tashiro}, {Townsley}, {Tueller},
  {Turner}, {Vietri}, {Voges}, {Ward}, {Willingale}, {Zerbi}, \&
  {Zhang}}]{2004ApJ...611.1005G}
{Gehrels}, N., {Chincarini}, G., {Giommi}, P., {et~al.}, 2004, \apj, 611, 1005

\bibitem[{{George} \& {Fabian}(1991)}]{1991MNRAS.249..352G}
{George}, I.~M. \& {Fabian}, A.~C., 1991, \mnras, 249, 352

\bibitem[{{Gliozzi} {et~al.}(2013){Gliozzi}, {Papadakis}, {Grupe}, {Brinkmann},
  \& {R{\"a}th}}]{2013MNRAS.433.1709G}
{Gliozzi}, M., {Papadakis}, I.~E., {Grupe}, D., {Brinkmann}, W.~P., \&
  {R{\"a}th}, C., 2013, \mnras, 433, 1709

\bibitem[{{Grupe} {et~al.}(2012){Grupe}, {Komossa}, {Gallo}, {Lia Longinotti},
  {Fabian}, {Pradhan}, {Gruberbauer}, \& {Xu}}]{2012ApJS..199...28G}
{Grupe}, D., {Komossa}, S., {Gallo}, L.~C., {Lia Longinotti}, A., {Fabian},
  A.~C., {Pradhan}, A.~K., {Gruberbauer}, M., \& {Xu}, D., 2012, \apjs, 199, 28

\bibitem[{{Haardt} \& {Maraschi}(1991)}]{1991ApJ...380L..51H}
{Haardt}, F. \& {Maraschi}, L., 1991, \apjl, 380, L51

\bibitem[{{Isobe} {et~al.}(1990){Isobe}, {Feigelson}, {Akritas}, \&
  {Babu}}]{1990ApJ...364..104I}
{Isobe}, T., {Feigelson}, E.~D., {Akritas}, M.~G., \& {Babu}, G.~J., 1990,
  \apj, 364, 104

\bibitem[{{Jansen} {et~al.}(2001){Jansen}, {Lumb}, {Altieri}, {Clavel}, {Ehle},
  {Erd}, {Gabriel}, {Guainazzi}, {Gondoin}, {Much}, {Munoz}, {Santos},
  {Schartel}, {Texier}, \& {Vacanti}}]{2001A&A...365L...1J}
{Jansen}, F., {Lumb}, D., {Altieri}, B., {et~al.}, 2001, \aap, 365, L1

\bibitem[{{Kara} {et~al.}(2013){Kara}, {Fabian}, {Cackett}, {Steiner},
  {Uttley}, {Wilkins}, \& {Zoghbi}}]{2013MNRAS.428.2795K}
{Kara}, E., {Fabian}, A.~C., {Cackett}, E.~M., {Steiner}, J.~F., {Uttley}, P.,
  {Wilkins}, D.~R., \& {Zoghbi}, A., 2013, \mnras, 428, 2795

\bibitem[{{Leighly}(1999)}]{1999ApJS..125..297L}
{Leighly}, K.~M., 1999, \apjs, 125, 297

\bibitem[{{Lubi{\'n}ski} {et~al.}(2016){Lubi{\'n}ski}, {Beckmann}, {Gibaud},
  {Paltani}, {Papadakis}, {Ricci}, {Soldi}, {T{\"u}rler}, {Walter}, \&
  {Zdziarski}}]{2016MNRAS.458.2454L}
{Lubi{\'n}ski}, P., {Beckmann}, V., {Gibaud}, L., {et~al.}, 2016, \mnras, 458,
  2454

\bibitem[{{Maiolino} \& {Risaliti}(2007)}]{2007ASPC..373..447M}
{Maiolino}, R. \& {Risaliti}, G., 2007, in Astronomical Society of the Pacific
  Conference Series, Vol. 373, The Central Engine of Active Galactic Nuclei,
  {Ho}, L.~C. \& {Wang}, J.-W., eds., p. 447

\bibitem[{{Malizia} {et~al.}(2014){Malizia}, {Molina}, {Bassani}, {Stephen},
  {Bazzano}, {Ubertini}, \& {Bird}}]{2014ApJ...782L..25M}
{Malizia}, A., {Molina}, M., {Bassani}, L., {Stephen}, J.~B., {Bazzano}, A.,
  {Ubertini}, P., \& {Bird}, A.~J., 2014, \apjl, 782, L25

\bibitem[{{Markowitz} {et~al.}(2009){Markowitz}, {Reeves}, {George}, {Braito},
  {Smith}, {Vaughan}, {Ar{\'e}valo}, \& {Tombesi}}]{2009ApJ...691..922M}
{Markowitz}, A., {Reeves}, J.~N., {George}, I.~M., {Braito}, V., {Smith}, R.,
  {Vaughan}, S., {Ar{\'e}valo}, P., \& {Tombesi}, F., 2009, \apj, 691, 922

\bibitem[{{Mason} {et~al.}(2001){Mason}, {Breeveld}, {Much}, {Carter},
  {Cordova}, {Cropper}, {Fordham}, {Huckle}, {Ho}, {Kawakami}, {Kennea},
  {Kennedy}, {Mittaz}, {Pandel}, {Priedhorsky}, {Sasseen}, {Shirey}, {Smith},
  \& {Vreux}}]{2001A&A...365L..36M}
{Mason}, K.~O., {Breeveld}, A., {Much}, R., {et~al.}, 2001, \aap, 365, L36

\bibitem[{{Mason} {et~al.}(2002){Mason}, {McHardy}, {Page}, {Uttley},
  {C{\'o}rdova}, {Maraschi}, {Priedhorsky}, {Puchnarewicz}, \&
  {Sasseen}}]{2002ApJ...580L.117M}
{Mason}, K.~O., {McHardy}, I.~M., {Page}, M.~J., {et~al.}, 2002, \apjl, 580,
  L117

\bibitem[{{McHardy} {et~al.}(2014){McHardy}, {Cameron}, {Dwelly}, {Connolly},
  {Lira}, {Emmanoulopoulos}, {Gelbord}, {Breedt}, {Arevalo}, \&
  {Uttley}}]{2014MNRAS.444.1469M}
{McHardy}, I.~M., {Cameron}, D.~T., {Dwelly}, T., {et~al.}, 2014, \mnras, 444,
  1469

\bibitem[{{Nandra} {et~al.}(1998){Nandra}, {Clavel}, {Edelson}, {George},
  {Malkan}, {Mushotzky}, {Peterson}, \& {Turner}}]{1998ApJ...505..594N}
{Nandra}, K., {Clavel}, J., {Edelson}, R.~A., {George}, I.~M., {Malkan}, M.~A.,
  {Mushotzky}, R.~F., {Peterson}, B.~M., \& {Turner}, T.~J., 1998, \apj, 505,
  594

\bibitem[{{Pan} {et~al.}(2016){Pan}, {Yuan}, {Yao}, {Zhou}, {Liu}, {Zhou}, \&
  {Zhang}}]{2016ApJ...819L..19P}
{Pan}, H.-W., {Yuan}, W., {Yao}, S., {Zhou}, X.-L., {Liu}, B., {Zhou}, H., \&
  {Zhang}, S.-N., 2016, \apjl, 819, L19

\bibitem[{{Peterson} {et~al.}(1998){Peterson}, {Wanders}, {Horne}, {Collier},
  {Alexander}, {Kaspi}, \& {Maoz}}]{1998PASP..110..660P}
{Peterson}, B.~M., {Wanders}, I., {Horne}, K., {Collier}, S., {Alexander}, T.,
  {Kaspi}, S., \& {Maoz}, D., 1998, \pasp, 110, 660

\bibitem[{{Petrucci} {et~al.}(2013){Petrucci}, {Paltani}, {Malzac}, {Kaastra},
  {Cappi}, {Ponti}, {De Marco}, {Kriss}, {Steenbrugge}, {Bianchi},
  {Branduardi-Raymont}, {Mehdipour}, {Costantini}, {Dadina}, \&
  {Lubi{\'n}ski}}]{2013A&A...549A..73P}
{Petrucci}, P.-O., {Paltani}, S., {Malzac}, J., {et~al.}, 2013, \aap, 549, A73

\bibitem[{{Remillard} {et~al.}(1986){Remillard}, {Bradt}, {Buckley}, {Roberts},
  {Schwartz}, {Tuohy}, \& {Wood}}]{1986ApJ...301..742R}
{Remillard}, R.~A., {Bradt}, H.~V., {Buckley}, D.~A.~H., {Roberts}, W.,
  {Schwartz}, D.~A., {Tuohy}, I.~R., \& {Wood}, K., 1986, \apj, 301, 742

\bibitem[{{Risaliti} {et~al.}(2005){Risaliti}, {Elvis}, {Fabbiano}, {Baldi}, \&
  {Zezas}}]{2005ApJ...623L..93R}
{Risaliti}, G., {Elvis}, M., {Fabbiano}, G., {Baldi}, A., \& {Zezas}, A., 2005,
  \apjl, 623, L93

\bibitem[{{Robertson} {et~al.}(2015){Robertson}, {Gallo}, {Zoghbi}, \&
  {Fabian}}]{2015MNRAS.453.3455R}
{Robertson}, D.~R.~S., {Gallo}, L.~C., {Zoghbi}, A., \& {Fabian}, A.~C., 2015,
  \mnras, 453, 3455

\bibitem[{{Romano} {et~al.}(2002){Romano}, {Turner}, {Mathur}, \&
  {George}}]{2002ApJ...564..162R}
{Romano}, P., {Turner}, T.~J., {Mathur}, S., \& {George}, I.~M., 2002, \apj,
  564, 162

\bibitem[{{Roming} {et~al.}(2005){Roming}, {Kennedy}, {Mason}, {Nousek}, {Ahr},
  {Bingham}, {Broos}, {Carter}, {Hancock}, {Huckle}, {Hunsberger}, {Kawakami},
  {Killough}, {Koch}, {McLelland}, {Smith}, {Smith}, {Soto}, {Boyd},
  {Breeveld}, {Holland}, {Ivanushkina}, {Pryzby}, {Still}, \&
  {Stock}}]{2005SSRv..120...95R}
{Roming}, P.~W.~A., {Kennedy}, T.~E., {Mason}, K.~O., {et~al.}, 2005, \ssr,
  120, 95

\bibitem[{{Ross} \& {Fabian}(2005)}]{2005MNRAS.358..211R}
{Ross}, R.~R. \& {Fabian}, A.~C., 2005, \mnras, 358, 211

\bibitem[{{Shapiro} {et~al.}(1976){Shapiro}, {Lightman}, \&
  {Eardley}}]{1976ApJ...204..187S}
{Shapiro}, S.~L., {Lightman}, A.~P., \& {Eardley}, D.~M., 1976, \apj, 204, 187

\bibitem[{{Shappee} {et~al.}(2014){Shappee}, {Prieto}, {Grupe}, {Kochanek},
  {Stanek}, {De Rosa}, {Mathur}, {Zu}, {Peterson}, {Pogge}, {Komossa}, {Im},
  {Jencson}, {Holoien}, {Basu}, {Beacom}, {Szczygie{\l}}, {Brimacombe},
  {Adams}, {Campillay}, {Choi}, {Contreras}, {Dietrich}, {Dubberley},
  {Elphick}, {Foale}, {Giustini}, {Gonzalez}, {Hawkins}, {Howell}, {Hsiao},
  {Koss}, {Leighly}, {Morrell}, {Mudd}, {Mullins}, {Nugent}, {Parrent},
  {Phillips}, {Pojmanski}, {Rosing}, {Ross}, {Sand}, {Terndrup}, {Valenti},
  {Walker}, \& {Yoon}}]{2014ApJ...788...48S}
{Shappee}, B.~J., {Prieto}, J.~L., {Grupe}, D., {et~al.}, 2014, \apj, 788, 48

\bibitem[{{Shemmer} {et~al.}(2001){Shemmer}, {Romano}, {Bertram}, {Brinkmann},
  {Collier}, {Crowley}, {Detsis}, {Filippenko}, {Gaskell}, {George}, {Gliozzi},
  {Hiller}, {Jewell}, {Kaspi}, {Klimek}, {Lannon}, {Li}, {Martini}, {Mathur},
  {Negoro}, {Netzer}, {Papadakis}, {Papamastorakis}, {Peterson}, {Peterson},
  {Pogge}, {Pronik}, {Rumstay}, {Sergeev}, {Sergeeva}, {Stirpe}, {Taylor},
  {Treffers}, {Turner}, {Uttley}, {Vestergaard}, {von Braun}, {Wagner}, \&
  {Zheng}}]{2001ApJ...561..162S}
{Shemmer}, O., {Romano}, P., {Bertram}, R., {et~al.}, 2001, \apj, 561, 162

\bibitem[{{Singh} {et~al.}(1985){Singh}, {Garmire}, \&
  {Nousek}}]{1985ApJ...297..633S}
{Singh}, K.~P., {Garmire}, G.~P., \& {Nousek}, J., 1985, \apj, 297, 633

\bibitem[{{Slone} \& {Netzer}(2012)}]{2012MNRAS.426..656S}
{Slone}, O. \& {Netzer}, H., 2012, \mnras, 426, 656

\bibitem[{{Sobolewska} \& {Done}(2005)}]{2005AIPC..774..317S}
{Sobolewska}, M. \& {Done}, C., 2005, in American Institute of Physics
  Conference Series, Vol. 774, X-ray Diagnostics of Astrophysical Plasmas:
  Theory, Experiment, and Observation, {Smith}, R., ed., pp. 317--319

\bibitem[{{Sobolewska} \& {Papadakis}(2009)}]{2009MNRAS.399.1597S}
{Sobolewska}, M.~A. \& {Papadakis}, I.~E., 2009, \mnras, 399, 1597

\bibitem[{{Str{\"u}der} {et~al.}(2001){Str{\"u}der}, {Briel}, {Dennerl},
  {Hartmann}, {Kendziorra}, {Meidinger}, {Pfeffermann}, {Reppin}, {Aschenbach},
  {Bornemann}, {Br{\"a}uninger}, {Burkert}, {Elender}, {Freyberg}, {Haberl},
  {Hartner}, {Heuschmann}, {Hippmann}, {Kastelic}, {Kemmer}, {Kettenring},
  {Kink}, {Krause}, {M{\"u}ller}, {Oppitz}, {Pietsch}, {Popp}, {Predehl},
  {Read}, {Stephan}, {St{\"o}tter}, {Tr{\"u}mper}, {Holl}, {Kemmer}, {Soltau},
  {St{\"o}tter}, {Weber}, {Weichert}, {von Zanthier}, {Carathanassis}, {Lutz},
  {Richter}, {Solc}, {B{\"o}ttcher}, {Kuster}, {Staubert}, {Abbey}, {Holland},
  {Turner}, {Balasini}, {Bignami}, {La Palombara}, {Villa}, {Buttler},
  {Gianini}, {Lain{\'e}}, {Lumb}, \& {Dhez}}]{2001A&A...365L..18S}
{Str{\"u}der}, L., {Briel}, U., {Dennerl}, K., {et~al.}, 2001, \aap, 365, L18

\bibitem[{{Sunyaev} \& {Titarchuk}(1980)}]{1980A&A....86..121S}
{Sunyaev}, R.~A. \& {Titarchuk}, L.~G., 1980, \aap, 86, 121

\bibitem[{{Turner} {et~al.}(1999){Turner}, {George}, {Nandra}, \&
  {Turcan}}]{1999ApJ...524..667T}
{Turner}, T.~J., {George}, I.~M., {Nandra}, K., \& {Turcan}, D., 1999, \apj,
  524, 667

\bibitem[{{Uttley}(2006)}]{2006ASPC..360..101U}
{Uttley}, P., 2006, in Astronomical Society of the Pacific Conference Series,
  Vol. 360, Astronomical Society of the Pacific Conference Series, {Gaskell},
  C.~M., {McHardy}, I.~M., {Peterson}, B.~M., \& {Sergeev}, S.~G., eds., p. 101

\bibitem[{{Vaughan} {et~al.}(2003){Vaughan}, {Edelson}, {Warwick}, \&
  {Uttley}}]{2003MNRAS.345.1271V}
{Vaughan}, S., {Edelson}, R., {Warwick}, R.~S., \& {Uttley}, P., 2003, \mnras,
  345, 1271

\bibitem[{{Wilkins} \& {Fabian}(2011)}]{2011MNRAS.414.1269W}
{Wilkins}, D.~R. \& {Fabian}, A.~C., 2011, \mnras, 414, 1269

\bibitem[{{Wilkins} \& {Fabian}(2012)}]{2012MNRAS.424.1284W}
---, 2012, \mnras, 424, 1284

\bibitem[{{Zdziarski}(1985)}]{1985ApJ...289..514Z}
{Zdziarski}, A.~A., 1985, \apj, 289, 514

\bibitem[{{Zdziarski} \& {Grandi}(2001)}]{2001ApJ...551..186Z}
{Zdziarski}, A.~A. \& {Grandi}, P., 2001, \apj, 551, 186

\end{thebibliography}
\label{lastpage}

\end{document}